\documentclass[prb,aps,twocolumn,amsmath,amssymb,superscriptaddress,floatfix,longbibliography]{revtex4-2}
\usepackage[breaklinks=true,colorlinks,citecolor=blue,linkcolor=blue,urlcolor=blue]{hyperref}
\usepackage{float}
\usepackage{bm}
\usepackage[normalem]{ulem}
\usepackage{epsfig,mathrsfs,color,latexsym,subfigure,marginnote,graphicx,verbatim,mathrsfs,color,array,amsmath,amsfonts,amssymb,graphicx}

\usepackage{multirow}
\usepackage{longtable}
\usepackage{relsize}

\usepackage[utf8]{inputenc}
\usepackage{physics}
\newcommand{\bk}{\vb{k}}
\newcommand{\bq}{\vb{q}}

\makeatletter
\def\maketitle{
\@author@finish
\title@column\titleblock@produce
\suppressfloats[t]}
\makeatother

\begin{document}
\title{Atomically thin obstructed atomic insulators with robust edge modes and quantized spin Hall effect}

\author{Rahul Verma}
\affiliation{Department of Condensed Matter Physics and Materials Science, Tata Institute of Fundamental Research, Colaba, Mumbai 400005, India}

\author{Shin-Ming Huang}
\affiliation{Department of Physics, National Sun Yat-sen University, Kaohsiung 80424, Taiwan}
\affiliation{Center for Theoretical and Computational Physics, National Sun Yat-sen University, Kaohsiung 80424, Taiwan}
\affiliation{Physics Division, National Center for Theoretical Sciences, Taipei 10617, Taiwan}

\author{ Bahadur Singh}
\email{Corresponding author: bahadur.singh@tifr.res.in}
\affiliation{Department of Condensed Matter Physics and Materials Science, Tata Institute of Fundamental Research, Colaba, Mumbai 400005, India}

\begin{abstract}
Symmetry-protected edge states serve as direct evidence of nontrivial electronic topology in atomically thin materials. Finding these states in experimentally realizable single-phase materials presents a substantial challenge for their use in fundamental studies and developing functional nanoscale devices. Here, we show the presence of robust edge states in phosphorene and group-Va monolayers with puckered lattice structures. By carefully analyzing the symmetry of the atomic sites and edge mode properties, we demonstrate that these atomically thin monolayers realize recently introduced obstructed atomic insulator states with partially occupied edge modes. The obstructed edge modes attain a Rashba-type spin splitting with Rashba parameter ($\alpha$) of 1.52 eV {\AA} for arsenene. Under strain or doping effects, these obstructed insulators transition to a phase with substantial spin-Berry curvature, yielding a double quantum spin Hall state with a spin Hall conductivity $\approx 4 \frac{e^2}{h}$. The experimental availability of phosphorene and other group-Va monolayers could enable verification of obstructed atomic states and enhanced spin-Berry curvature effects discussed in this study, offering the potential for applications in topological electronic and spintronic devices. 
\end{abstract}

\maketitle
\section{Introduction}
Coupling of electron topology, spin, and orbital degrees of freedom is the foundation of topological quantum materials having desirable disorder-resistant states~\cite{RMP2010,QMat2017,Singh2022}. The nontrivial topological state is generally driven by a bulk band inversion at high symmetry points in the Brillouin zone (BZ) and thus the information of electronic wavefunction symmetries at these high-symmetry points is enough to diagnose the topology of a material as proposed in theory of topological quantum chemistry or symmetry indicators~\cite{PRX_2017, SI_ashvin, Song2018,TQC_Bradlyn}. Especially, within topological quantum chemistry, the band representations (BRs) of occupied states of a material can be expressed as a linear combination of the elementary band representations (eBRs, also known as the basis of atomic limit) at high-symmetry points. When specific coefficients of this linear combination of eBRs are rational fractions, the material is a topological insulator or topological semimetal. Conversely, if these coefficients are non-negative integers, the material is a trivial insulator~\cite{TQC_Bradlyn, EBs_2001, EBRs_2018, PRL_2018}. The topologically trivial insulators can also support unconventional surface states if their electrons occupy atom-unoccupied Wyckoff positions such that the BRs cannot be induced only from atom-centered band representations (aBRs)~\cite{FeOAI, Uncov_mat,OAI_RSI}. Such insulators, named obstructed atomic insulators (OAIs), host electronic states with filling anomaly at surfaces cleaved at atom-unoccupied Wyckoff positions~\cite{OAI_RSI}. Notably, the band representations of electrides, which form ionic crystals where electrons detach and occupy interstitial or vacancy sites, cannot be decomposed as a sum of aBR. Consequently, while they may exhibit edge states similar to those in OAIs, additional criteria must still be met for OAIs to form electrides~\cite{TQC_electride}.The OAIs have been predicted in three-dimensional (3D) materials and verified recently in SrIn$_2$P$_2$ and elemental silicon~\cite{OAI_SrIn2P2, OAI_Si}. The OAIs hold great promise for realizing exotic states such as higher-order topological states, unconventional superconductivity, and electrides with superior catalytic properties useful for quantum science and technological applications~\cite{PRB2022_Li, OAI_cornermodes, PRB_corner_charge,OAI_MS2Z4, OAI_majorana, OAI_superconductivity, OAI_QSH, OAI_active_sites, OAI_mott}. 
 
The atomically thin 2D materials provide an unprecedented framework for investigating topological quantum phenomena under carrier confinement that can be further manipulated by external strain, electric field, or doping effects~\cite{2DvdW,Rasbha_As,phosphorene,P_isoelectron}. Among various families of 2D materials, phosphorene and group-Va materials with puckered honeycomb lattice are mechanically exfoliated to the monolayer limit~\cite{phosphorene, P_isoelectron, phos_science, PRB2021_Sb, alpha_Bi,Rajibul_MSi2Z4,Unpinned_Sb}. They exhibit a thickness-dependent bandgap that is further amenable to external electric field and surface doping effects. Specifically, a topologically nontrivial Dirac semimetal state can be realized in phosphorene through potassium surface doping or the application of a vertical electric field~\cite{Ghosh2016, phos_doping, phos_Efield, phos_pressure}. The other group-Va monolayer layers as one moves down the Periodic Table (P$\rightarrow$As$\rightarrow$Sb) show a topological phase transition to a state with multiple unpinned Dirac cones and puzzling intertwining edge states~\cite{Unpinned_Sb, Unpinned_Dirac, HSCI_BaoKai}. Despite these studies, it is unclear if these 2D materials can support OAIs with conducting edge modes and how the edge modes evolve across the Periodic Table in these materials.

In this paper, we report the theoretical prediction of the OAI state with spin-filtered edge modes in monolayers of group-Va materials with puckered honeycomb lattice. Our systematic analysis of the bulk and edge energy spectra combined with symmetry analysis illustrates that phosphorene and group-Va monolayers exhibit half-filled obstructed states along armchair and zigzag edges. The obstructed edge states (OESs) exhibit a finite Rashba-type spin splitting, attaining a Rashba coupling constant of $\alpha=1.52$ eV {\AA} for arsenene. We discuss the evolution of OESs from P to As and Sb and show that due to a generic point band inversion in Sb, these states evolve to form double quantum spin Hall edge states in it with a spin-Hall conductivity of $\sigma_{xy}^z \approx 4\frac{e^2}{h}$. Such topological phase transition can be captured by applying uniaxial strain in arsenene. We further demonstrate that these materials respect spin U(1) quasisymmetry, which ascertains a nearly quantized spin Hall conductivity and two pairs of spin-polarized counterpropagating topological edge states. 

Quasisymmetry refers to an approximate symmetry of the Hamiltonian that remains exact under first-order perturbations but breaks down in higher-order terms~\cite{QuasiSymm,QuasiSymm_GT}. This can lead to a small hybridization gap between states of different characters, involving distinct degrees of freedom such as valley, orbital, or spin. Notably, a perturbation of strength $\lambda$ can hybridize these states; its effects are typically not noticeable until the second- or higher-order terms of $\lambda$. When $\lambda$ varies with energy (or $k$), the hybridization gap remains negligible at low energy but becomes significant at higher energies. The spin U(1) quasisymmetry is a feature that $S_z$ is preserved at low energies with weak higher-order spin-mixing corrections~\cite{QuasiSpin,Quasi_ESCI}. To demonstrate the spin U(1) quasisymmetry in these monolayers, we develop a minimal $k \cdot p$ model Hamiltonian that elucidates the nature of calculated spin-filtered edge states. Our findings indicate that group-Va monolayers are promising atomically thin materials for exploring OAIs, spin U(1) quasisymmetry, and nearly quantized spin Hall effect.

\section{Crystal structure and OAI state} 
\begin{figure*}[t!]
\centering
\includegraphics[width=0.80\textwidth]{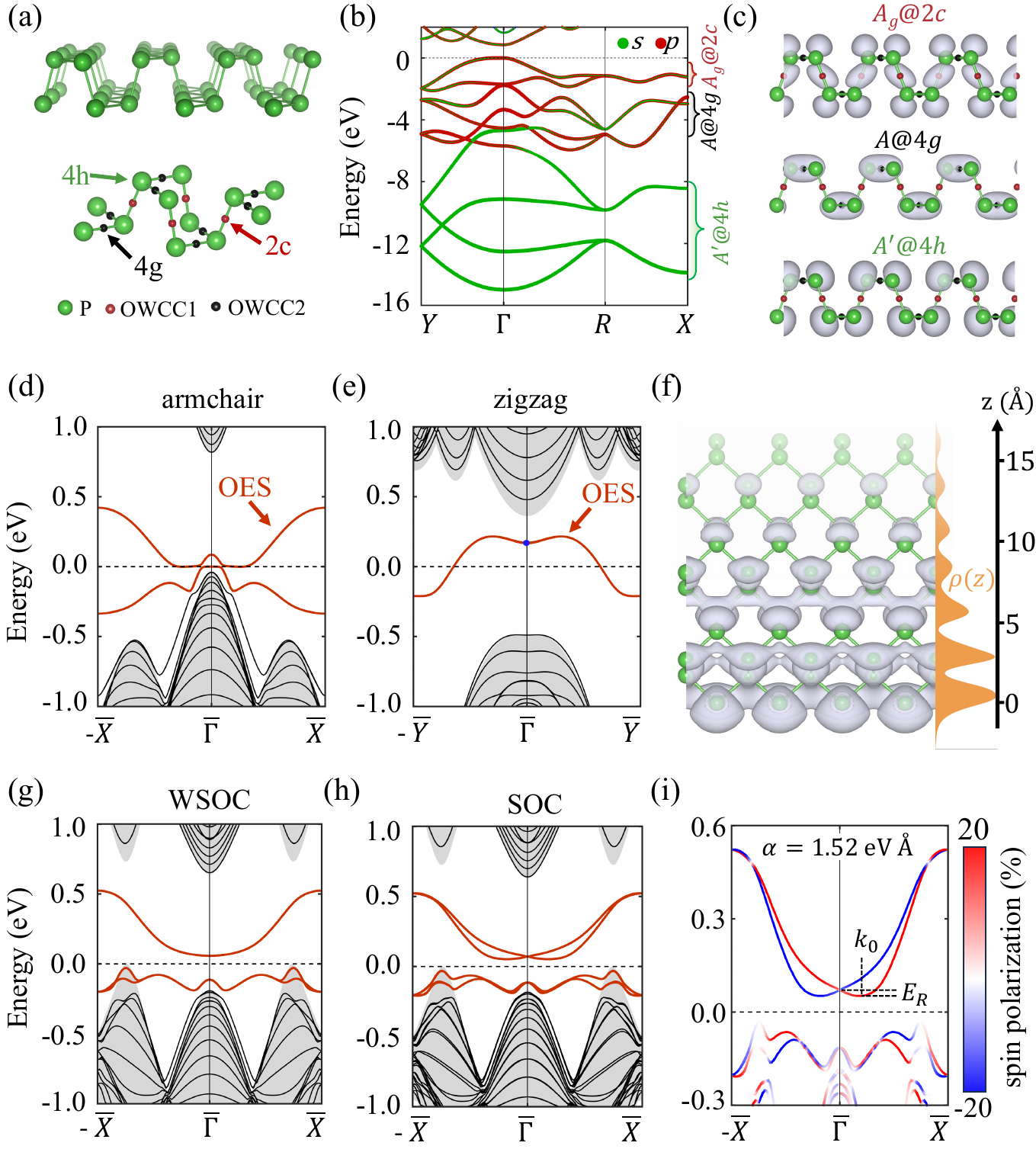}
\caption{(a) Puckered honeycomb lattice of phosphorene with Wyckoff positions $4h$ (green), $2c$ (red), and $4g$ (black). $4h$ Wyckoff position is atom occupied, whereas $2c$ and $4g$ positions are unoccupied (OWCC). (b) Orbital resolved band structure of phosphorene without spin-orbit coupling (SOC). Green and red colors identify P $s$ and $p$ states, respectively. BRs of occupied bands are shown. (c) Electronic charge density at $X$ point arising from the BRs of $A_g@2c$, $A@4g$, and $A^\prime@4h$ (see text for details). The maximum charge for $A_g@2c$ and $A@4g$ BRs is centered at atom unoccupied positions $2c$ and $4g$. (d),(e) Edge spectrum of phosphorene for (d) armchair and (e) zigzag edges without SOC. The shaded gray region identifies projected bulk bands and red lines mark the OESs. (f) Electronic charge density ($\rho$) of OESs at the $\overline{\Gamma}$ point for zigzag edge [blue circle in (e)]. The right panel shows the variation of $\rho$ as a function of slab thickness $z$ in orange. The vertical distance is calculated from the bottom phosphorus atom. Band structure of arsenene (g) without and (h) with SOC along the armchair edge. (i) Spin texture of OESs along the armchair edge for arsenene. Blue and red colors indicate up and down spin channels of $S_z$. $k_0$ and $E_R$ denote momentum and energy offsets of spin-split states.}
 \label{OAI}
\end{figure*}

We begin by discussing the puckered honeycomb structure of phosphorene, which constitutes a single layer of bulk black phosphorus. This structure consists of two stacked zigzag chains of phosphorus atoms connected by out-of-plane bonds, with each atom forming three bonds with its nearest neighbors [Fig.~\ref{OAI}(a)]. This particular bonding configuration leads to a non-symmorphic space group, $Pmna$ (No. 53). It respects inversion $\mathcal{I}:(x,y,z)\rightarrow (-x,-y,-z)$, a vertical mirror plane $\mathcal{M}_y:(x,y,z)\rightarrow (x,-y,z)$ perpendicular to the zigzag direction and twofold rotations $\mathcal{\widetilde{C}}_{2x}:(x,y,z)\rightarrow (x+\frac{1}{2},-y+\frac{1}{2},-z)$,  $\mathcal{C}_{2y}:(x,y,z)\rightarrow (-x,y,-z)$, $\mathcal{\widetilde{C}}_{2z}:(x,y,z)\rightarrow (-x+\frac{1}{2},-y+\frac{1}{2},z)$, and glide mirror $\mathcal{\widetilde{M}}_{x}:(x,y,z)\rightarrow (-x+\frac{1}{2},y+\frac{1}{2}, z)$ and $\mathcal{\widetilde{M}}_{z}:(x,y,z)\rightarrow (x+\frac{1}{2},y+\frac{1}{2},-z)$ symmetries. P atoms occupy $4h: \{(0, y, z), (0,-y, -z), (\frac{1}{2},-y, z+ \frac{1}{2}), (\frac{1}{2}, y, -z+ \frac{1}{2})  \}$ Wyckoff position in the lattice that induces 4 (8) BRs of single (double) space group. There are two additional Wyckoff positions $2c: \{(\frac{1}{2}, \frac{1}{2}, 0), (0, \frac{1}{2}, \frac{1}{2})\}$ and $4g: \{\ (\frac{1}{4}, y, \frac{1}{4}), (\frac{1}{4}, -y, \frac{3}{4}), (\frac{3}{4}, -y, \frac{3}{4}), (\frac{3}{4}, y, \frac{1}{4}) \}$ in lattice that are not occupied by atoms. Considering five valence electrons of P and any group-Va element, the phosphorene lattice has 20 valence electrons. This satisfies a condition for filling-enforced OAI, $N_e=8\mathbb{N}+4$, where $\mathbb{N} \in \mathbb{Z}$,  for space group P$mna$~\cite{FeOAI}. This implies that the decomposition of BR into eBRs should include BRs generated from atom-unoccupied Wyckoff positions.

To identify the BRs of unoccupied Wyckoff positions, we present the band structure of phosphorene in Fig.~\ref{OAI}(b) [see Supplemental Material (SM) for calculational details~\cite{supp}]. The valence and conduction bands are well separated with an energy gap, dictating an insulator ground state. A mapping of irreducible representations (IRs) of occupied bands by considering the eBRs of the single-space group reveals that $s$ bands belong to BRs induced from atom-occupied Wyckoff position $4h$ [marked as $A'@4h$ in Fig.~\ref{OAI}(b)]. The remaining $p$-type bands belong to BRs that are generated from atom-unoccupied Wyckoff positions $2c$ and $4g$ ($A_g@2c$ and $A@4g$). This is more resolved in real-space charge localization of band representations in Fig.~\ref{OAI}(c). Specifically, the electronic charge density distribution arising from BRs of $A_g@2c$ and $A@4g$ at $X$ point are localized on the empty Wyckoff positions $2c$ and $4g$, respectively. In contrast, the charge density of $A^\prime@4h$ states is centered at atom-occupied Wyckoff position $4h$. The bands with $p$ orbital character in phosphorene thus arise from unoccupied or obstructed atomic sites in the lattice.

We further evaluate the single-valued real space invariant (RSI) $\delta_{1}$ at Wyckoff positions $2c$, $4g$, and $4h$ of space group $Pmna$ to elucidate the OAI state~\cite{OAI_RSI}. The RSI index at each Wyckoff position is defined as $\delta_1@2c = -m(A_g)+m(A_u)-m(B_g)+m(B_u)$, $\delta_1@4g =-m(A)+m(B)$, and $\delta_1@4h = -m(A^\prime)+m(A^{\prime \prime})$, where $m(\rho)$ is the number of eBRs ($\rho$) induced from a Wyckoff position. We obtain $\delta_1@2c=-1$, $\delta_1@4g=-1$, and $\delta_1@4h=-2$. The nonzero integer values of RSI at $2c$ and $4g$ confirm the OAI phase in phosphorene. A similar analysis holds for all other group-Va monolayers with a puckered lattice. The nontrivial RSI index indicates that these monolayers will support robust states over edges that cut through these unoccupied Wyckoff positions.

In Figs.~\ref{OAI}(d) and \ref{OAI}(e), we present the first-principles edge band structure of phosphorene terminated at $2c$ (zigzag edge) and $4g$ (armchair edge) sites without SOC. Both edges support the partially filled conducting states within the 2D bulk energy gap. Notably, the OESs do not connect the bulk valence and conduction bands and thus are distinct from quantum spin Hall insulators where the helical edge states connect the bulk valence and conduction bands. Regardless, the OESs cross the Fermi level, exhibiting the filling anomaly~\cite{FeOAI, Uncov_mat,OAI_RSI}. Figure~\ref{OAI}(f) illustrates the distribution of the electronic charge density associated with OESs at $\overline{\Gamma}$ point for the zigzag edge and its variation as a function of slab thickness ($z$). The electronic charge density is localized at the slab edge and decays exponentially into the bulk as expected for edge states. We further present the obstructed edge states (OESs) of arsenene for the armchair edge in Figs.~\ref{OAI}(g)$-$\ref{OAI}(i). Two slightly gapped OESs are observed near the Fermi level in the absence of spin-orbit coupling (SOC). With the inclusion of SOC, these OESs exhibit Rashba-type spin splitting with a maximum energy offset of $E_R=0.019$ eV at $k_0=0.025$ Å$^{-1}$ and a Rashba constant of $\alpha=\frac{2E_R}{k_0}=1.52$ eVÅ. Note that the OESs in phosphorene also show spin splitting; however, due to the small SOC in phosphorus, the magnitude of this spin splitting is relatively small (see SM~\cite{supp}).
\begin{figure*}[t!]
\centering
\includegraphics[width=0.80\textwidth]{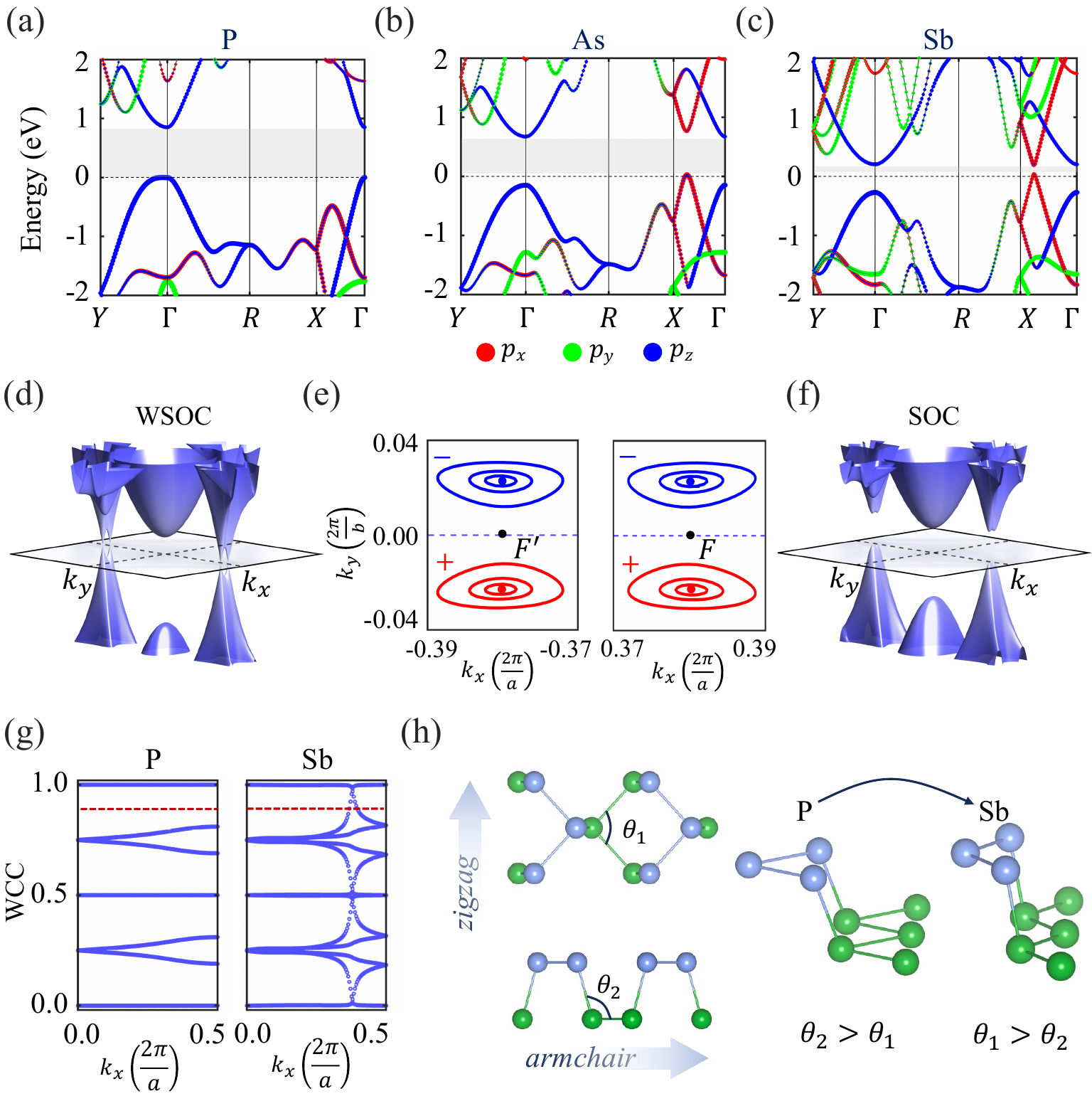}
\caption{Band structure of monolayer (a) P,  (b) As, and (c) Sb with puckered honeycomb lattice without SOC. Red, green, and blue identify $p_x$, $p_y$, and $p_z$ orbitals. Bulk energy gap is marked in shaded gray. (d) $E-k_x-k_y$ energy dispersion and (e) constant energy contours of Sb without SOC. Two pairs of Dirac cones are resolved at generic $k$ points. $F$ or $F'$ mark the midpoint of each pair on the $\Gamma-X$ line. The symbols $\pm$ denote the winding number (chirality) of the Dirac nodes. (f) $E-k_x-k_y$ energy dispersion of Sb with SOC showing a hybridization gap at Dirac points. (g) WCC spectrum for P and Sb. WCCs cross the arbitrary red dashed line an even number of times for Sb. (h) Evolution of crystal lattice from P to Sb with increased puckering. Green and blue spheres identify atoms on two different planes. $\theta_1$ and $\theta_2$ define in-plane and out-of-plane angles (left). Two geometric configurations exposing hidden structural transition from P to Sb (right).}
\label{band_topology}
\end{figure*}

\section{Hidden structural and topological phase transition} 

\begin{figure*}[t]
\includegraphics[width=0.80\textwidth]{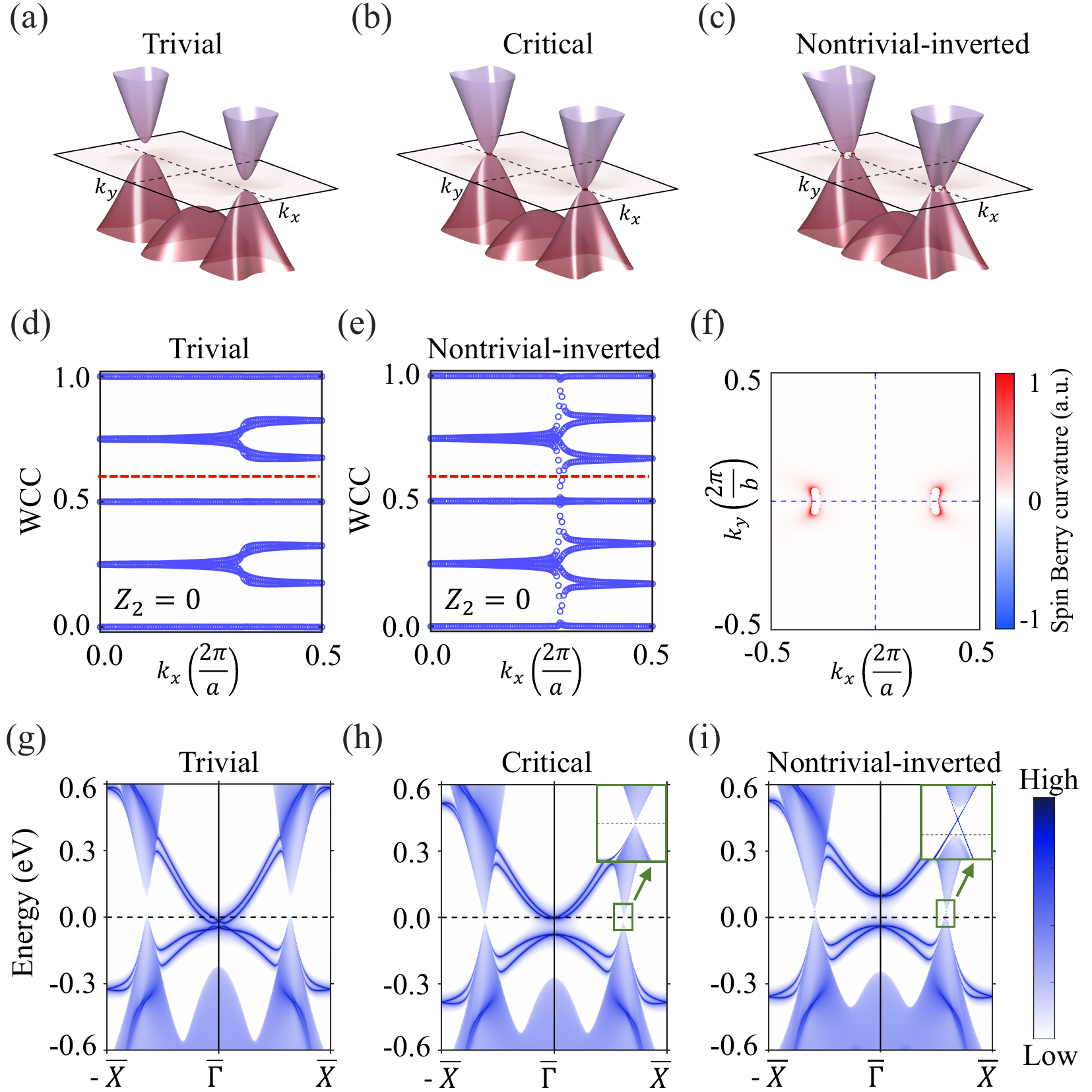}
\caption{Calculated $E-k_x-k_y$ band structure of strained As without SOC for (a) $b^\prime=1.10 b$, trivial phase, (b) $b^\prime=1.12 b$, critical phase, and (c) $b^\prime=1.14 b$, nontrivial-inverted phase. $b$ is the lattice parameter along the zigzag direction. Dashed lines on the plane mark $k_x$ and $k_y$ principal axes of 2D BZ. Evolution of WCCs for (d) $b^\prime=1.10 b$; trivial phase and (e) $b^\prime=1.14 b$; nontrivial-inverted phase with SOC. (f) Spin-Berry curvature obtained for $b^\prime=1.14 b$, showing hotspots at the hybridized Dirac nodes. Calculated semi-infinite edge spectrum of As for (g) $b^\prime=1.10 b$, (h) $b^\prime=1.12 b$, and (i) $b^\prime=1.14 b$  with SOC. OESs evolve from gapped to gapless on the $\overline{\Gamma}-\overline{X}$ line.}
\label{phase_transition}
\end{figure*}

We now discuss the evolution of electronic structure from monolayer P to Sb and reveal a unique topological phase transition. In Figs.~\ref{band_topology}(a)$-$\ref{band_topology}(c), we present the orbital-resolved band structures of these monolayers in the vicinity of the Fermi level without SOC. Although the band structure stays insulating from P to Sb along the high-symmetry directions, there are subtle changes in it. The valence and conduction band extremum is shifted from the $\Gamma$ point in P to a generic momentum point in Sb. The bands consist of $p_z$ orbital at the $\Gamma$ point whereas they are comprised of $p_x$ orbitals along the $\Gamma-X$ direction. These $p_x$ bands change dispersion from parabolic in P and As to almost linear and gapless in Sb. Notably, although a small gap is present along the $\Gamma-X$ line in Sb, these bands exhibit gapless behavior away from this line, resulting in the formation of four Dirac cones in the $k_x-k_y$ plane at generic momenta $(\pm 0.38\frac{2\pi}{a}, \pm 0.02 \frac{2\pi}{b})$ [see Fig.~\ref{band_topology}(e)]. Given the sublattice symmetry of the puckered lattice, which is ensured by the $\mathcal{I}$, $\mathcal{C}_{2y}$, or $\mathcal{C}_{2z}$ symmetries, these Dirac cones are stable from gap opening in the absence of SOC~\cite{Unpinned_Sb}. The sublattice symmetry allows the definition of a winding number along the closed loop around the Dirac nodes as $W_n=\frac{1}{\pi}\oint A_k.d\bm{k}$ where, $A_k$ is the Berry connection of the occupied bands. Our explicit numerical calculations confirm the nontrivial winding number of $\pm 1$ as marked in Fig.~\ref{band_topology}(e). In the presence of SOC, a small hybridization gap opens up at the Dirac nodes, leading to an inverted band structure with band inversion at generic $k$ points [Fig.~\ref{band_topology}(f)]. 

\begin{figure*}[t]
\centering
\includegraphics[width=0.80\textwidth]{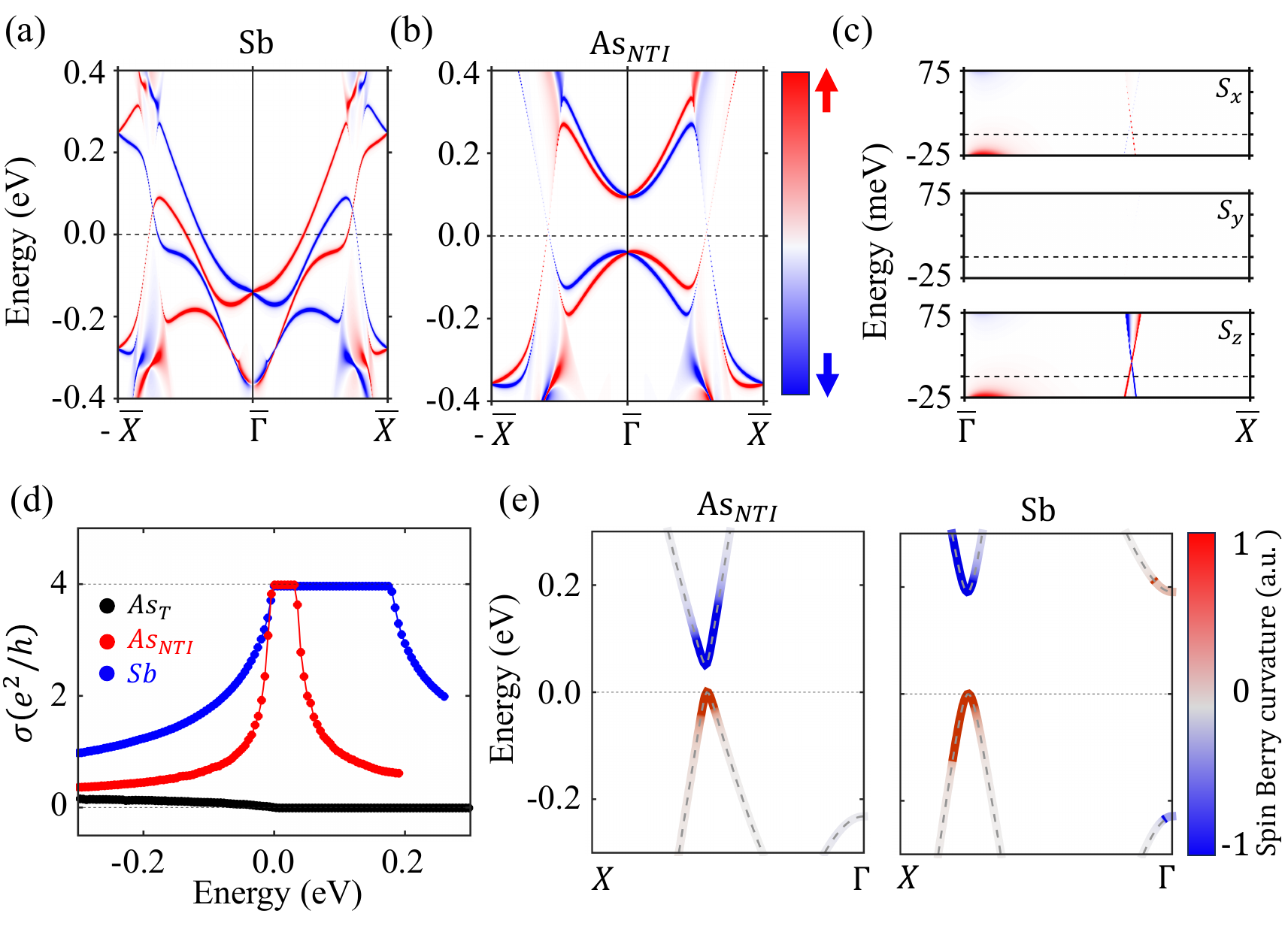}
\caption{$S_z$ spin-component resolved edge spectrum of (a) Sb and (b) nontrivial-inverted As (As$_{NTI}$) with $b^\prime=1.14b$. Red and blue indicate spin-up and spin-down states polarized along the $z$ direction. (c) Spin components resolved edge spectrum of As$_{NTI}$ with $b^\prime=1.14b$. Spin is nearly polarized along the $z$ direction, maintaining spin U(1) quasisymmetry. (d) Intrinsic SHC $\sigma_{xy}^{z}$ for trivial and nontrivial phases of As and Sb. SHC is nearly quantized to 4$\frac{e^2}{h}$ within the bulk energy gap. (e) Band resolved spin-Berry curvature along the $\Gamma-X$ direction for As$_{NTI}$ and Sb.}
 \label{spin}
\end{figure*}

In Fig.~\ref{band_topology}(g), we present the Wannier charge centers (WCCs) spectrum of the occupied states of P and Sb. The WCCs spectrum shows adiabatically distinct characteristics, remaining gapful for P and gapless for Sb. Despite these features, the WCCs spectrum reveals $\mathbb{Z}_2$ trivial connectivity for both P and Sb, thereby dictating a $\mathbb{Z}_2=0$ trivial state. The topological $\mathbb{Z}_2$ trivial state is further verified through the parity analysis of occupied bands at high-symmetry points. The presence of a band inversion and the WCCs connectivity in Sb signals a nontrivial electronic state distinct from P. However, since the band inversion happens at generic $k$ points, it does not alter IRs of occupied bands at the high-symmetry points. Such a topological state is hidden from symmetry-indicator and TQC theories and, therefore, P, As, and Sb share the same topological characterization based on these theories~\cite{Unpinned_Dirac,HSCI_BaoKai}.

We emphasize that the emergence of electronic structure from P to Sb has a strong correlation with their puckered lattice geometry and signals a hidden structural transition. To facilitate this analysis, we define bond angles $\theta_1$ and $\theta_2$ in the puckered lattice as shown in Fig.~\ref{band_topology}(h). $\theta_1$ is an in-plane angle in the zigzag layer and $\theta_2$ is the out-of-plane angle along the armchair direction, describing the nonplanner puckering of the lattice. $\theta_1$ and $\theta_2$ are $95.97^{\circ}$ and $103.87^{\circ}$ for P and $96.04^{\circ}$ and $95.67^{\circ}$ for Sb, respectively. This indicates that the in-plane bond angle $\theta_1$ is smaller than the nonplanner bond angle $\theta_2$ ($\theta_1 < \theta_2$) for P with less puckering. This bond angle order is reversed for Sb with $\theta_1 > \theta_2$ with large puckering. The electronic phase transition happens when $\theta_1 \sim \theta_2$. Importantly, when $\theta_1 \sim \theta_2$, each atom shares an isotropic electronic environment similar to graphene, thereby forming two critical points on the $\Gamma-X$ line without SOC. When $\theta_1 > \theta_2$, each critical point splits into two Dirac points away from the $\Gamma-X$ line as seen in Sb.

A similar topological phase transition can also happen in As under the application of uniaxial strain along the zigzag direction. In Figs.~\ref{phase_transition}(a)$-$\ref{phase_transition}(c), we present the calculated band structure of As at various values of strain. We change the lattice parameter $b$ along the zigzag direction (see SM for details~\cite{supp}). When $b^\prime=1.10 b$, the valence and conduction bands have a parabolic energy dispersion with minimum along the $\Gamma-X$ direction. With an increase in $b^\prime$, the valence and conduction bands start approaching each other and cross at the $\Gamma-X$ line for $b^\prime=1.12 b$ [Fig.~\ref{phase_transition}(b)]. With a further increase in $b^\prime$, four Dirac cones appear at the generic $k$ points across the $\Gamma-X$ line. In Figs.~\ref{phase_transition}(d) and \ref{phase_transition}(e), we present WCCs spectrum of As for $b^\prime = 1.10 b$ and $b^\prime = 1.14b$ with SOC. The WCCs spectrum resolves distinct characteristics with a $\mathbb{Z}_2=0$ trivial phase. However, a strong spin-Berry curvature is found near the gapped Dirac nodes for As with generic $k$ point band inversion [Fig.~\ref{phase_transition}(f)]. 

To explore the emergence of OESs with strain, we present the calculated armchair edge spectrum in Figs.~\ref{phase_transition}(g)$-$\ref{phase_transition}(i) for $b^\prime = 1.10 b$, $b^\prime = 1.12b$, and  $b^\prime = 1.14b$ with SOC. The two pairs of OESs are seen within the bulk gap in the vicinity of the Fermi level for $b^\prime = 1.10 b$ [Fig.~\ref{phase_transition}(g)]. These OESs evolve to form gapless Dirac-type states within the bulk band inversion at generic $k$ points, forming double quantum spin-Hall edge states.

\section{Spin Hall effect and model Hamiltonian} 
The preceding discussion made it clear that, although Sb and strained As are classified as OAIs based on the analysis of IRs at high-symmetry points, they possess a topologically distinct band structure compared to P or As. Their OESs form a gapless Dirac-type energy dispersion within the bulk gap. In Figs.~\ref{spin}(a) and \ref{spin}(b), we present the spin texture of these edge states for Sb and As with $b^\prime=1.14 b$. The two pairs of oppositely spin-polarized edge states are clearly observed along the $-\overline{X}-\overline{\Gamma}-\overline{X}$ direction, with a dominant $S_z$ spin component near the edge Dirac nodes. 
Specifically, the presence of time-reversal and $C_{2y}$ symmetries along the armchair edge permits both $S_x$ and $S_z$ spin components for the edge states. However, near the edge Dirac nodes, $S_x$ is found to be negligible, leaving only the $S_z$ component [see Fig.~\ref{spin}(c)]. This particular spin texture allows the definition of the spin U(1) quasisymmetry with an $S_z$-preserved spin Hamiltonian, enabling the system to support a nearly quantized spin-Hall conductivity.

Figure~\ref{spin}(d) illustrates the calculated intrinsic spin-Hall conductivity (SHC) $\sigma_{xy}^{z}$ as a function of the Fermi energy for As and Sb. The SHC is calculated using the Kubo formula~\cite{shc_TaAs, SHC_w90} 
\begin{equation}
\sigma_{xy}^{z}=\frac{-e^2}{\hbar} \frac{1}{A} \sum_k \Omega_{xy}^z (\bm{k}) 
\end{equation}
Here 
\begin{equation}
\Omega_{xy}^z(\bm{k})= \sum_n f_{n} (\bm{k})\Omega_{n,xy}^z(\bm{k})
\end{equation}
is the $k$-resolved spin Berry curvature and 
\begin{equation}
\Omega_{n,xy}^z(\bm{k})= {\hbar^2} \sum_{m\ne n} \frac {-2\text{Im}\langle n\bm{k}|  \hat{J_x^z} |m\bm{k}\rangle \langle m\bm{k}|  \hat v_y |n\bm{k}\rangle}{(E_{nk}-E_{mk})^2}
\end{equation}
 gives the band resolved spin Berry curvature in 2D BZ with area $A$.  $E_{nk}$ represents the energy of Bloch state $|n\bm{k}\rangle$ with occupation $f_{n}(\bm{k})$. The spin current operator is defined as $\hat{J_x^z}=\frac{1}{2}\{\hat\sigma_z, \hat v_x\}$, where $\hat\sigma_z$ is the spin operator and $\hat v_x$ is the velocity operator. $\sigma_{xy}^{z}$ represents $z$-polarized spin-current along $x$ direction generated by an electric field along $y$ direction. The SHC for the As is zero, whereas it has a value of  $\approx 3.96 \frac{e^2}{h}$ for Sb and $\approx 3.98 \frac{e^2}{h}$ for strained As ($b^\prime>1.12 b$) 
 in the bulk gap region. Such a nearly quantized large SHC arises mainly due to strong spin-Berry curvature at the band inverted $k$ points in both Sb and As$_{NTI}$ [Fig. ~\ref{spin}(e)]. 

Generally, the SHC deviates from exact quantization in materials due to nonconserved-$S_z$ Hamiltonians~\cite{Kane_mele_QSH,PRL2006,Pordan_2009,PRL_2005}. However, we now show that these group-Va monolayers host a unique SOC Hamiltonian that conserves $S_z$ spin component, maintaining spin U(1) quasisymmetry and a nearly quantized spin Hall effect. Since the spin-Berry curvature is primarily concentrated around the generic band inversion points, we discuss a model Hamiltonian around these points [$F~(F')$ points on the $\overline{\Gamma X}$ line; see Fig.~\ref{band_topology}(e)]. Notably, Ref.~\cite{Unpinned_Dirac} provides a detailed discussion of the model Hamiltonian near the $F~(F')$ points and we refer to their results with additional discussion on quasisymmetry.

The puckered hexagonal unit cell of group-Va monolayers contains four atoms, named A$_U$, A$_L$, B$_U$, and B$_L$, where U and L refer to the upper and lower planes, respectively (see SM~\cite{supp}). Due to the identical nature of all atoms in the unit cell, the structure exhibits chiral (sublattice) symmetry. The $k \cdot p$ model in the absence of SOC at $F$ is thus a two-band model, where each band is characterized by a sublattice degree of freedom. In the basis of the chiral operator $\mathcal{C} = \sigma_z$, the inversion ($\mathcal{I}$) and rotational symmetries interchange sublattices via the operation $\sigma_x$. When spin ($s$) is considered, these symmetry operators evolve as $\mathcal{I}=\sigma_x s_0$, $\mathcal{C}_{2y}=-i\sigma_x s_y$, $\tilde{\mathcal{C}}_{2z} = -i\sigma_x s_z$, and the time-reversal operator $\mathcal{T} = -i s_y K$, where $s_0 $ and $s_{x,y,z}$ are the identity and Pauli matrices in spin space. Given the symmetry constraints imposed by these operators, the spin-ful $k \cdot p$ model Hamiltonian is expressed as ~\cite{Unpinned_Dirac} 
\begin{equation}
	H_{\mu}(\bq) = d_x(\bq) \sigma_x  s_0 +  d_y(\bq) \sigma_y  s_0 + \Delta q_y \sigma_z s_z,
\end{equation}
with $d_x(\bq) = \mu a_1 q_x + a_2 q_x^2 + a_3(q_y^2 - h^2)$ and $ d_y(\bq) = b_1 q_x + \mu \left( b_2 q_x^2 +  b_3 (q_y^2 - h^2) \right)$, where $\mu = \pm $ corresponds to the $F$ and $F'$ points, respectively, and $\bq$ is measured from $F$ (or $F'$). $a$'s and $b$'s are the expansion coefficients. In the absence of SOC ($\Delta=0$), two Dirac nodes emerge near $F$ (or $F'$) at $q_y = \pm h$, whereas the SOC introduces a gap at these Dirac nodes of magnitude $\abs{\Delta h}$. Importantly, the SOC comes with $s_z$. The Hamiltonian commutes with $s_z$ indicating that $S_z$ is a good quantum number for states near $F$ and $F'$ points.

We also consider the model Hamiltonian near the $\Gamma$ point. As the low-energy bands near the $\Gamma$ point originate from the $p_z$ orbitals, we consider one orbital per sublattice with only the nearest neighbor hoppings to derive the model Hamiltonian~\cite{Unpinned_Dirac}. Considering $\sigma$ acting on A and B sublattices and $\tau$ on upper and lower planes, the symmetry operators are written as $\mathcal{I}=\sigma_x \tau_x$, $\mathcal{C}_{2y}=-i \sigma_x \tau_x s_y$, $\tilde{\mathcal{C}}_{2z} = -i\sigma_x s_z$, and $\mathcal{T} = -i s_y K$. The corresponding spinless Hamiltonian is given by,
\begin{equation}
	H_{\Gamma}^{t}(\bk) =  t_{\perp} \sigma_x \tau_x + \text{Re} (Q_{11}) \sigma_x \tau_0 - \text{Im} (Q_{11}) \sigma_y \tau_0,
\end{equation}
where $Q_{11} = 2 t_1 e^{k_x/2} \cos \frac{k_y}{2} + 2 t_2 e^{-k_x/2} \cos \frac{k_y}{2} \approx 2(t_1+t_2) + i(t_1 - t_2)k_x + \mathcal{O}(\bk^2) $.  $t_{1}$ and $t_{2}$ are in-plane nearest-neighbor hoppings for the bottom and top plane, respectively, and $t_{\perp}$ is out-of-plane hopping (see SM~\cite{supp}). Here $\tau_x$ is a good quantum number with $\tau_x = \pm 1$. The low-energy two bands correspond to $\tau_x=-1$, with energies $E^{\tau_x=-1}_{\pm}(\bk) = \pm \sqrt{[t_{\perp}-2(t_1+t_2)]^2+(t_1-t_2)^2k_x^2}$. The symmetry allowed SOC Hamiltonian is written as, 
\begin{align}
	\begin{split}
	H_{\Gamma}^{\text{SOC}}(\bk) = & \sigma_0 \tau_z (A_1 k_y s_x + A_2 k_x s_y) \\
	&+ \sigma_x \tau_z (B_1 k_y s_x + B_2 k_x s_y) \\
	&+ C \sigma_y \tau_z s_y + D \sigma_z \tau_0 s_z,
	\end{split}
\end{align}
where $A$'s, $B$'s, $C$, and $D$ are material dependent parameters. This SOC Hamiltonian does not preserve the $S_z$ symmetry near the $\Gamma$ point. However, in the eigenbasis of $H_{\Gamma}^{t}(\bk)$ with $\tau_x=-1$, the SOC Hamiltonian is approximated to
\begin{equation}
	H_{\Gamma}^{\text{SOC}}(\bk) \approx  D' k_y \sigma_z \tau_0 s_z
\end{equation}
since $\matrixel{\alpha, \tau_x=-1}{\tau_z}{\beta, \tau_x=-1} =0$ (this is the same for $\tau_x=+1$). As a result, $S_z$ is conserved around $\Gamma$. We emphasize that the conservation of $S_z$ describes the spin U(1) quasisymmetry, which is an approximate symmetry to the first order in SOC. The second-order perturbation hybridizes $\tau_x = \pm1$ states and,  consequently, the spin-up and spin-down states may hybridize. However, the quasisymmetry remains a robust feature in these materials, as the bands with $\tau_x = -1$ and $\tau_x = +1$ are well separated, with an energy gap much larger than the SOC strength.

Since $S_z$ is conserved both around the $F~(F')$ and $\Gamma$ points, the spin eigenstates can be decomposed into spin-up ($s_z=+1$) and spin-down ($s_z=-1$) sectors, reflecting spin U(1) symmetry in these monolayers. The spin current, defined as the difference between the currents of opposite spins, is influenced by the large spin Berry curvature, which is concentrated near the $F~(F')$ points for Sb and strained nontrivial-inverted As. This results in a nearly quantized spin-Hall conductivity. Due to the presence of an $S_z$-preserving SOC Hamiltonian with spin U(1) quasisymmetry, the associated topological state can be characterized by a spin-Chern number of 2~\cite{PRL2006,Pordan_2009,PRL_2005,QuasiSpin}.

\section{Summary} 
In summary, we have demonstrated the existence of an OAI state in phosphorene and related group-Va monolayers with a puckered lattice structure. The charge density of their occupied bands near the Fermi level is centered at atom-unoccupied Wyckoff positions, leading to a nontrivial RSI index and the emergence of OESs on crystal edges that cut through these unoccupied Wyckoff positions. We show that these OESs are spin polarized, exhibiting a large Rashba-type spin splitting with a Rashba parameter ($\alpha$) of 1.52 eV·Å for As. Additionally, we resolve a unique topological state with band inversion at generic $k$ points in Sb and strained As and nearly quantized SHC of $4\frac{e^2}{h}$. Based on the spin-polarized edge states and model Hamiltonian analysis, we show that the nearly quantization of SHC is attributed to a unique $S_z$ conserved SOC Hamiltonian that maintains spin U(1) quasi-symmetry. Our work thus demonstrates that phosphorene and other group-Va monolayers are promising candidates for exploring OAIs and enhanced spin-Berry curvature effects, with potential applications in topological spintronics.

\section{Acknowledgements} We thank B. Patra for the useful discussions. This work is supported by the Department of Atomic Energy of the Government of India under Project No. 12-R$\&$D-TFR-5.10-0100 and benefited from the HPC resources of TIFR Mumbai. S.M.H. is supported by the NSTC-AFOSR Taiwan program on Topological and Nanostructured Materials, Grant No. 110-2124-M-110-002-MY3.
 
\bibliography{Obstructed_HSI}

\begin{thebibliography}{62}%
\makeatletter
\providecommand \@ifxundefined [1]{%
 \@ifx{#1\undefined}
}%
\providecommand \@ifnum [1]{%
 \ifnum #1\expandafter \@firstoftwo
 \else \expandafter \@secondoftwo
 \fi
}%
\providecommand \@ifx [1]{%
 \ifx #1\expandafter \@firstoftwo
 \else \expandafter \@secondoftwo
 \fi
}%
\providecommand \natexlab [1]{#1}%
\providecommand \enquote  [1]{``#1''}%
\providecommand \bibnamefont  [1]{#1}%
\providecommand \bibfnamefont [1]{#1}%
\providecommand \citenamefont [1]{#1}%
\providecommand \href@noop [0]{\@secondoftwo}%
\providecommand \href [0]{\begingroup \@sanitize@url \@href}%
\providecommand \@href[1]{\@@startlink{#1}\@@href}%
\providecommand \@@href[1]{\endgroup#1\@@endlink}%
\providecommand \@sanitize@url [0]{\catcode `\\12\catcode `\$12\catcode
  `\&12\catcode `\#12\catcode `\^12\catcode `\_12\catcode `\%12\relax}%
\providecommand \@@startlink[1]{}%
\providecommand \@@endlink[0]{}%
\providecommand \url  [0]{\begingroup\@sanitize@url \@url }%
\providecommand \@url [1]{\endgroup\@href {#1}{\urlprefix }}%
\providecommand \urlprefix  [0]{URL }%
\providecommand \Eprint [0]{\href }%
\providecommand \doibase [0]{https://doi.org/}%
\providecommand \selectlanguage [0]{\@gobble}%
\providecommand \bibinfo  [0]{\@secondoftwo}%
\providecommand \bibfield  [0]{\@secondoftwo}%
\providecommand \translation [1]{[#1]}%
\providecommand \BibitemOpen [0]{}%
\providecommand \bibitemStop [0]{}%
\providecommand \bibitemNoStop [0]{.\EOS\space}%
\providecommand \EOS [0]{\spacefactor3000\relax}%
\providecommand \BibitemShut  [1]{\csname bibitem#1\endcsname}%
\let\auto@bib@innerbib\@empty
\bibitem [{\citenamefont {Hasan}\ and\ \citenamefont {Kane}(2010)}]{RMP2010}%
  \BibitemOpen
  \bibfield  {author} {\bibinfo {author} {\bibfnamefont {M.~Z.}\ \bibnamefont
  {Hasan}}\ and\ \bibinfo {author} {\bibfnamefont {C.~L.}\ \bibnamefont
  {Kane}},\ }\bibfield  {title} {\bibinfo {title} {Colloquium: Topological
  insulators},\ }\href {https://doi.org/10.1103/RevModPhys.82.3045} {\bibfield
  {journal} {\bibinfo  {journal} {Rev. Mod. Phys.}\ }\textbf {\bibinfo {volume}
  {82}},\ \bibinfo {pages} {3045} (\bibinfo {year} {2010})}\BibitemShut
  {NoStop}%
\bibitem [{\citenamefont {Tokura}\ \emph {et~al.}(2017)\citenamefont {Tokura},
  \citenamefont {Kawasaki},\ and\ \citenamefont {Nagaosa}}]{QMat2017}%
  \BibitemOpen
  \bibfield  {author} {\bibinfo {author} {\bibfnamefont {Y.}~\bibnamefont
  {Tokura}}, \bibinfo {author} {\bibfnamefont {M.}~\bibnamefont {Kawasaki}},\
  and\ \bibinfo {author} {\bibfnamefont {N.}~\bibnamefont {Nagaosa}},\
  }\bibfield  {title} {\bibinfo {title} {Emergent functions of quantum
  materials},\ }\href {https://doi.org/10.1038/nphys4274} {\bibfield  {journal}
  {\bibinfo  {journal} {Nat. Phys.}\ }\textbf {\bibinfo {volume} {13}},\
  \bibinfo {pages} {1056} (\bibinfo {year} {2017})}\BibitemShut {NoStop}%
\bibitem [{\citenamefont {Singh}\ \emph {et~al.}(2022)\citenamefont {Singh},
  \citenamefont {Lin},\ and\ \citenamefont {Bansil}}]{Singh2022}%
  \BibitemOpen
  \bibfield  {author} {\bibinfo {author} {\bibfnamefont {B.}~\bibnamefont
  {Singh}}, \bibinfo {author} {\bibfnamefont {H.}~\bibnamefont {Lin}},\ and\
  \bibinfo {author} {\bibfnamefont {A.}~\bibnamefont {Bansil}},\ }\bibfield
  {title} {\bibinfo {title} {Topology and symmetry in quantum materials},\
  }\href {https://doi.org/10.1002/adma.202201058} {\bibfield  {journal}
  {\bibinfo  {journal} {Adv. Mater.}\ }\textbf {\bibinfo {volume} {35}},\
  \bibinfo {pages} {2201058} (\bibinfo {year} {2022})}\BibitemShut {NoStop}%
\bibitem [{\citenamefont {Kruthoff}\ \emph {et~al.}(2017)\citenamefont
  {Kruthoff}, \citenamefont {de~Boer}, \citenamefont {van Wezel}, \citenamefont
  {Kane},\ and\ \citenamefont {Slager}}]{PRX_2017}%
  \BibitemOpen
  \bibfield  {author} {\bibinfo {author} {\bibfnamefont {J.}~\bibnamefont
  {Kruthoff}}, \bibinfo {author} {\bibfnamefont {J.}~\bibnamefont {de~Boer}},
  \bibinfo {author} {\bibfnamefont {J.}~\bibnamefont {van Wezel}}, \bibinfo
  {author} {\bibfnamefont {C.~L.}\ \bibnamefont {Kane}},\ and\ \bibinfo
  {author} {\bibfnamefont {R.-J.}\ \bibnamefont {Slager}},\ }\bibfield  {title}
  {\bibinfo {title} {Topological classification of crystalline insulators
  through band structure combinatorics},\ }\href
  {https://doi.org/10.1103/PhysRevX.7.041069} {\bibfield  {journal} {\bibinfo
  {journal} {Phys. Rev. X}\ }\textbf {\bibinfo {volume} {7}},\ \bibinfo {pages}
  {041069} (\bibinfo {year} {2017})}\BibitemShut {NoStop}%
\bibitem [{\citenamefont {Po}\ \emph {et~al.}(2017)\citenamefont {Po},
  \citenamefont {Vishwanath},\ and\ \citenamefont {Watanabe}}]{SI_ashvin}%
  \BibitemOpen
  \bibfield  {author} {\bibinfo {author} {\bibfnamefont {H.~C.}\ \bibnamefont
  {Po}}, \bibinfo {author} {\bibfnamefont {A.}~\bibnamefont {Vishwanath}},\
  and\ \bibinfo {author} {\bibfnamefont {H.}~\bibnamefont {Watanabe}},\
  }\bibfield  {title} {\bibinfo {title} {Symmetry-based indicators of band
  topology in the 230 space groups},\ }\href
  {https://doi.org/10.1038/s41467-017-00133-2} {\bibfield  {journal} {\bibinfo
  {journal} {Nat. Commun.}\ }\textbf {\bibinfo {volume} {8}},\ \bibinfo {pages}
  {50} (\bibinfo {year} {2017})}\BibitemShut {NoStop}%
\bibitem [{\citenamefont {Song}\ \emph {et~al.}(2018)\citenamefont {Song},
  \citenamefont {Zhang}, \citenamefont {Fang},\ and\ \citenamefont
  {Fang}}]{Song2018}%
  \BibitemOpen
  \bibfield  {author} {\bibinfo {author} {\bibfnamefont {Z.}~\bibnamefont
  {Song}}, \bibinfo {author} {\bibfnamefont {T.}~\bibnamefont {Zhang}},
  \bibinfo {author} {\bibfnamefont {Z.}~\bibnamefont {Fang}},\ and\ \bibinfo
  {author} {\bibfnamefont {C.}~\bibnamefont {Fang}},\ }\bibfield  {title}
  {\bibinfo {title} {Quantitative mappings between symmetry and topology in
  solids},\ }\href {https://doi.org/10.1038/s41467-018-06010-w} {\bibfield
  {journal} {\bibinfo  {journal} {Nat. Commun.}\ }\textbf {\bibinfo {volume}
  {9}},\ \bibinfo {pages} {3530} (\bibinfo {year} {2018})}\BibitemShut
  {NoStop}%
\bibitem [{\citenamefont {Bradlyn}\ \emph {et~al.}(2017)\citenamefont
  {Bradlyn}, \citenamefont {Elcoro}, \citenamefont {Cano}, \citenamefont
  {Vergniory}, \citenamefont {Wang}, \citenamefont {Felser}, \citenamefont
  {Aroyo},\ and\ \citenamefont {Bernevig}}]{TQC_Bradlyn}%
  \BibitemOpen
  \bibfield  {author} {\bibinfo {author} {\bibfnamefont {B.}~\bibnamefont
  {Bradlyn}}, \bibinfo {author} {\bibfnamefont {L.}~\bibnamefont {Elcoro}},
  \bibinfo {author} {\bibfnamefont {J.}~\bibnamefont {Cano}}, \bibinfo {author}
  {\bibfnamefont {M.~G.}\ \bibnamefont {Vergniory}}, \bibinfo {author}
  {\bibfnamefont {Z.}~\bibnamefont {Wang}}, \bibinfo {author} {\bibfnamefont
  {C.}~\bibnamefont {Felser}}, \bibinfo {author} {\bibfnamefont {M.~I.}\
  \bibnamefont {Aroyo}},\ and\ \bibinfo {author} {\bibfnamefont {B.~A.}\
  \bibnamefont {Bernevig}},\ }\bibfield  {title} {\bibinfo {title} {Topological
  quantum chemistry},\ }\href {https://doi.org/10.1038/nature23268} {\bibfield
  {journal} {\bibinfo  {journal} {Nature}\ }\textbf {\bibinfo {volume} {547}},\
  \bibinfo {pages} {298} (\bibinfo {year} {2017})}\BibitemShut {NoStop}%
\bibitem [{\citenamefont {Michel}(2001)}]{EBs_2001}%
  \BibitemOpen
  \bibfield  {author} {\bibinfo {author} {\bibfnamefont {L.}~\bibnamefont
  {Michel}},\ }\bibfield  {title} {\bibinfo {title} {Elementary energy bands in
  crystals are connected},\ }\href
  {https://doi.org/10.1016/s0370-1573(00)00093-4} {\bibfield  {journal}
  {\bibinfo  {journal} {Phys. Rep.}\ }\textbf {\bibinfo {volume} {341}},\
  \bibinfo {pages} {377} (\bibinfo {year} {2001})}\BibitemShut {NoStop}%
\bibitem [{\citenamefont {Cano}\ \emph
  {et~al.}(2018{\natexlab{a}})\citenamefont {Cano}, \citenamefont {Bradlyn},
  \citenamefont {Wang}, \citenamefont {Elcoro}, \citenamefont {Vergniory},
  \citenamefont {Felser}, \citenamefont {Aroyo},\ and\ \citenamefont
  {Bernevig}}]{EBRs_2018}%
  \BibitemOpen
  \bibfield  {author} {\bibinfo {author} {\bibfnamefont {J.}~\bibnamefont
  {Cano}}, \bibinfo {author} {\bibfnamefont {B.}~\bibnamefont {Bradlyn}},
  \bibinfo {author} {\bibfnamefont {Z.}~\bibnamefont {Wang}}, \bibinfo {author}
  {\bibfnamefont {L.}~\bibnamefont {Elcoro}}, \bibinfo {author} {\bibfnamefont
  {M.~G.}\ \bibnamefont {Vergniory}}, \bibinfo {author} {\bibfnamefont
  {C.}~\bibnamefont {Felser}}, \bibinfo {author} {\bibfnamefont {M.~I.}\
  \bibnamefont {Aroyo}},\ and\ \bibinfo {author} {\bibfnamefont {B.~A.}\
  \bibnamefont {Bernevig}},\ }\bibfield  {title} {\bibinfo {title} {Building
  blocks of topological quantum chemistry: Elementary band representations},\
  }\href {https://doi.org/10.1103/PhysRevB.97.035139} {\bibfield  {journal}
  {\bibinfo  {journal} {Phys. Rev. B}\ }\textbf {\bibinfo {volume} {97}},\
  \bibinfo {pages} {035139} (\bibinfo {year} {2018}{\natexlab{a}})}\BibitemShut
  {NoStop}%
\bibitem [{\citenamefont {Cano}\ \emph
  {et~al.}(2018{\natexlab{b}})\citenamefont {Cano}, \citenamefont {Bradlyn},
  \citenamefont {Wang}, \citenamefont {Elcoro}, \citenamefont {Vergniory},
  \citenamefont {Felser}, \citenamefont {Aroyo},\ and\ \citenamefont
  {Bernevig}}]{PRL_2018}%
  \BibitemOpen
  \bibfield  {author} {\bibinfo {author} {\bibfnamefont {J.}~\bibnamefont
  {Cano}}, \bibinfo {author} {\bibfnamefont {B.}~\bibnamefont {Bradlyn}},
  \bibinfo {author} {\bibfnamefont {Z.}~\bibnamefont {Wang}}, \bibinfo {author}
  {\bibfnamefont {L.}~\bibnamefont {Elcoro}}, \bibinfo {author} {\bibfnamefont
  {M.~G.}\ \bibnamefont {Vergniory}}, \bibinfo {author} {\bibfnamefont
  {C.}~\bibnamefont {Felser}}, \bibinfo {author} {\bibfnamefont {M.~I.}\
  \bibnamefont {Aroyo}},\ and\ \bibinfo {author} {\bibfnamefont {B.~A.}\
  \bibnamefont {Bernevig}},\ }\bibfield  {title} {\bibinfo {title} {Topology of
  disconnected elementary band representations},\ }\href
  {https://doi.org/10.1103/PhysRevLett.120.266401} {\bibfield  {journal}
  {\bibinfo  {journal} {Phys. Rev. Lett.}\ }\textbf {\bibinfo {volume} {120}},\
  \bibinfo {pages} {266401} (\bibinfo {year} {2018}{\natexlab{b}})}\BibitemShut
  {NoStop}%
\bibitem [{\citenamefont {Xu}\ \emph {et~al.}(2024)\citenamefont {Xu},
  \citenamefont {Elcoro}, \citenamefont {Song}, \citenamefont {Vergniory},
  \citenamefont {Felser}, \citenamefont {Parkin}, \citenamefont {Regnault},
  \citenamefont {Ma\~nes},\ and\ \citenamefont {Bernevig}}]{FeOAI}%
  \BibitemOpen
  \bibfield  {author} {\bibinfo {author} {\bibfnamefont {Y.}~\bibnamefont
  {Xu}}, \bibinfo {author} {\bibfnamefont {L.}~\bibnamefont {Elcoro}}, \bibinfo
  {author} {\bibfnamefont {Z.-D.}\ \bibnamefont {Song}}, \bibinfo {author}
  {\bibfnamefont {M.~G.}\ \bibnamefont {Vergniory}}, \bibinfo {author}
  {\bibfnamefont {C.}~\bibnamefont {Felser}}, \bibinfo {author} {\bibfnamefont
  {S.~S.~P.}\ \bibnamefont {Parkin}}, \bibinfo {author} {\bibfnamefont
  {N.}~\bibnamefont {Regnault}}, \bibinfo {author} {\bibfnamefont {J.~L.}\
  \bibnamefont {Ma\~nes}},\ and\ \bibinfo {author} {\bibfnamefont {B.~A.}\
  \bibnamefont {Bernevig}},\ }\bibfield  {title} {\bibinfo {title}
  {Filling-enforced obstructed atomic insulators},\ }\href
  {https://doi.org/10.1103/PhysRevB.109.165139} {\bibfield  {journal} {\bibinfo
   {journal} {Phys. Rev. B}\ }\textbf {\bibinfo {volume} {109}},\ \bibinfo
  {pages} {165139} (\bibinfo {year} {2024})}\BibitemShut {NoStop}%
\bibitem [{\citenamefont {Gao}\ \emph {et~al.}(2022)\citenamefont {Gao},
  \citenamefont {Qian}, \citenamefont {Jia}, \citenamefont {Guo}, \citenamefont
  {Fang}, \citenamefont {Liu}, \citenamefont {Weng},\ and\ \citenamefont
  {Wang}}]{Uncov_mat}%
  \BibitemOpen
  \bibfield  {author} {\bibinfo {author} {\bibfnamefont {J.}~\bibnamefont
  {Gao}}, \bibinfo {author} {\bibfnamefont {Y.}~\bibnamefont {Qian}}, \bibinfo
  {author} {\bibfnamefont {H.}~\bibnamefont {Jia}}, \bibinfo {author}
  {\bibfnamefont {Z.}~\bibnamefont {Guo}}, \bibinfo {author} {\bibfnamefont
  {Z.}~\bibnamefont {Fang}}, \bibinfo {author} {\bibfnamefont {M.}~\bibnamefont
  {Liu}}, \bibinfo {author} {\bibfnamefont {H.}~\bibnamefont {Weng}},\ and\
  \bibinfo {author} {\bibfnamefont {Z.}~\bibnamefont {Wang}},\ }\bibfield
  {title} {\bibinfo {title} {Unconventional materials: the mismatch between
  electronic charge centers and atomic positions},\ }\href
  {https://doi.org/10.1016/j.scib.2021.12.025} {\bibfield  {journal} {\bibinfo
  {journal} {Sci. Bull.}\ }\textbf {\bibinfo {volume} {67}},\ \bibinfo {pages}
  {598} (\bibinfo {year} {2022})}\BibitemShut {NoStop}%
\bibitem [{\citenamefont {Xu}\ \emph {et~al.}(2021)\citenamefont {Xu},
  \citenamefont {Elcoro}, \citenamefont {Li}, \citenamefont {Song},
  \citenamefont {Regnault}, \citenamefont {Yang}, \citenamefont {Sun},
  \citenamefont {Parkin}, \citenamefont {Felser},\ and\ \citenamefont
  {Bernevig}}]{OAI_RSI}%
  \BibitemOpen
  \bibfield  {author} {\bibinfo {author} {\bibfnamefont {Y.}~\bibnamefont
  {Xu}}, \bibinfo {author} {\bibfnamefont {L.}~\bibnamefont {Elcoro}}, \bibinfo
  {author} {\bibfnamefont {G.}~\bibnamefont {Li}}, \bibinfo {author}
  {\bibfnamefont {Z.-D.}\ \bibnamefont {Song}}, \bibinfo {author}
  {\bibfnamefont {N.}~\bibnamefont {Regnault}}, \bibinfo {author}
  {\bibfnamefont {Q.}~\bibnamefont {Yang}}, \bibinfo {author} {\bibfnamefont
  {Y.}~\bibnamefont {Sun}}, \bibinfo {author} {\bibfnamefont {S.}~\bibnamefont
  {Parkin}}, \bibinfo {author} {\bibfnamefont {C.}~\bibnamefont {Felser}},\
  and\ \bibinfo {author} {\bibfnamefont {B.~A.}\ \bibnamefont {Bernevig}},\
  }\bibfield  {title} {\bibinfo {title} {Three-dimensional real space
  invariants, obstructed atomic insulators and a new principle for active
  catalytic sites},\ }\href {https://arxiv.org/abs/2111.02433} {\  (\bibinfo
  {year} {2021})},\ \Eprint {https://arxiv.org/abs/2111.02433}
  {arXiv:2111.02433} \BibitemShut {NoStop}%
\bibitem [{\citenamefont {Nie}\ \emph {et~al.}(2021)\citenamefont {Nie},
  \citenamefont {Qian}, \citenamefont {Gao}, \citenamefont {Fang},
  \citenamefont {Weng},\ and\ \citenamefont {Wang}}]{TQC_electride}%
  \BibitemOpen
  \bibfield  {author} {\bibinfo {author} {\bibfnamefont {S.}~\bibnamefont
  {Nie}}, \bibinfo {author} {\bibfnamefont {Y.}~\bibnamefont {Qian}}, \bibinfo
  {author} {\bibfnamefont {J.}~\bibnamefont {Gao}}, \bibinfo {author}
  {\bibfnamefont {Z.}~\bibnamefont {Fang}}, \bibinfo {author} {\bibfnamefont
  {H.}~\bibnamefont {Weng}},\ and\ \bibinfo {author} {\bibfnamefont
  {Z.}~\bibnamefont {Wang}},\ }\bibfield  {title} {\bibinfo {title}
  {Application of topological quantum chemistry in electrides},\ }\href
  {https://doi.org/10.1103/PhysRevB.103.205133} {\bibfield  {journal} {\bibinfo
   {journal} {Phys. Rev. B}\ }\textbf {\bibinfo {volume} {103}},\ \bibinfo
  {pages} {205133} (\bibinfo {year} {2021})}\BibitemShut {NoStop}%
\bibitem [{\citenamefont {Liu}\ \emph {et~al.}(2023)\citenamefont {Liu},
  \citenamefont {Deng}, \citenamefont {Liu}, \citenamefont {Yin}, \citenamefont
  {Chen}, \citenamefont {Zhu}, \citenamefont {Yang}, \citenamefont {Jiang},
  \citenamefont {Liu}, \citenamefont {Ye}, \citenamefont {Shen}, \citenamefont
  {Yin}, \citenamefont {Wang}, \citenamefont {Liu}, \citenamefont {Zhao},\ and\
  \citenamefont {Liu}}]{OAI_SrIn2P2}%
  \BibitemOpen
  \bibfield  {author} {\bibinfo {author} {\bibfnamefont {X.-R.}\ \bibnamefont
  {Liu}}, \bibinfo {author} {\bibfnamefont {H.}~\bibnamefont {Deng}}, \bibinfo
  {author} {\bibfnamefont {Y.}~\bibnamefont {Liu}}, \bibinfo {author}
  {\bibfnamefont {Z.}~\bibnamefont {Yin}}, \bibinfo {author} {\bibfnamefont
  {C.}~\bibnamefont {Chen}}, \bibinfo {author} {\bibfnamefont {Y.-P.}\
  \bibnamefont {Zhu}}, \bibinfo {author} {\bibfnamefont {Y.}~\bibnamefont
  {Yang}}, \bibinfo {author} {\bibfnamefont {Z.}~\bibnamefont {Jiang}},
  \bibinfo {author} {\bibfnamefont {Z.}~\bibnamefont {Liu}}, \bibinfo {author}
  {\bibfnamefont {M.}~\bibnamefont {Ye}}, \bibinfo {author} {\bibfnamefont
  {D.}~\bibnamefont {Shen}}, \bibinfo {author} {\bibfnamefont {J.-X.}\
  \bibnamefont {Yin}}, \bibinfo {author} {\bibfnamefont {K.}~\bibnamefont
  {Wang}}, \bibinfo {author} {\bibfnamefont {Q.}~\bibnamefont {Liu}}, \bibinfo
  {author} {\bibfnamefont {Y.}~\bibnamefont {Zhao}},\ and\ \bibinfo {author}
  {\bibfnamefont {C.}~\bibnamefont {Liu}},\ }\bibfield  {title} {\bibinfo
  {title} {Spectroscopic signature of obstructed surface states in
  {SrIn}$_{2}${P}$_{2}$},\ }\href {https://doi.org/10.1038/s41467-023-38589-0}
  {\bibfield  {journal} {\bibinfo  {journal} {Nat. Commun.}\ }\textbf {\bibinfo
  {volume} {14}},\ \bibinfo {pages} {2905} (\bibinfo {year}
  {2023})}\BibitemShut {NoStop}%
\bibitem [{\citenamefont {Liu}\ \emph {et~al.}(2024{\natexlab{a}})\citenamefont
  {Liu}, \citenamefont {Deng}, \citenamefont {Xu}, \citenamefont {Yang},
  \citenamefont {Pei}, \citenamefont {Chen}, \citenamefont {He}, \citenamefont
  {Liu}, \citenamefont {Mo}, \citenamefont {Kim}, \citenamefont {Cacho},
  \citenamefont {Yao}, \citenamefont {Song}, \citenamefont {Chen},
  \citenamefont {Wang}, \citenamefont {Yan}, \citenamefont {Yang},
  \citenamefont {Bernevig},\ and\ \citenamefont {Chen}}]{OAI_Si}%
  \BibitemOpen
  \bibfield  {author} {\bibinfo {author} {\bibfnamefont {Z.}~\bibnamefont
  {Liu}}, \bibinfo {author} {\bibfnamefont {P.}~\bibnamefont {Deng}}, \bibinfo
  {author} {\bibfnamefont {Y.}~\bibnamefont {Xu}}, \bibinfo {author}
  {\bibfnamefont {H.}~\bibnamefont {Yang}}, \bibinfo {author} {\bibfnamefont
  {D.}~\bibnamefont {Pei}}, \bibinfo {author} {\bibfnamefont {C.}~\bibnamefont
  {Chen}}, \bibinfo {author} {\bibfnamefont {S.}~\bibnamefont {He}}, \bibinfo
  {author} {\bibfnamefont {D.}~\bibnamefont {Liu}}, \bibinfo {author}
  {\bibfnamefont {S.-K.}\ \bibnamefont {Mo}}, \bibinfo {author} {\bibfnamefont
  {T.}~\bibnamefont {Kim}}, \bibinfo {author} {\bibfnamefont {C.}~\bibnamefont
  {Cacho}}, \bibinfo {author} {\bibfnamefont {H.}~\bibnamefont {Yao}}, \bibinfo
  {author} {\bibfnamefont {Z.-D.}\ \bibnamefont {Song}}, \bibinfo {author}
  {\bibfnamefont {X.}~\bibnamefont {Chen}}, \bibinfo {author} {\bibfnamefont
  {Z.}~\bibnamefont {Wang}}, \bibinfo {author} {\bibfnamefont {B.}~\bibnamefont
  {Yan}}, \bibinfo {author} {\bibfnamefont {L.}~\bibnamefont {Yang}}, \bibinfo
  {author} {\bibfnamefont {B.~A.}\ \bibnamefont {Bernevig}},\ and\ \bibinfo
  {author} {\bibfnamefont {Y.}~\bibnamefont {Chen}},\ }\bibfield  {title}
  {\bibinfo {title} {Massive 1d dirac line, solitons and reversible
  manipulation on the surface of a prototype obstructed atomic insulator,
  silicon},\ }\href {https://arxiv.org/abs/2406.08114} {\  (\bibinfo {year}
  {2024}{\natexlab{a}})},\ \Eprint {https://arxiv.org/abs/2406.08114}
  {arXiv:2406.08114} \BibitemShut {NoStop}%
\bibitem [{\citenamefont {Li}\ \emph {et~al.}(2022)\citenamefont {Li},
  \citenamefont {Ma}, \citenamefont {Liu}, \citenamefont {Yu},\ and\
  \citenamefont {Yao}}]{PRB2022_Li}%
  \BibitemOpen
  \bibfield  {author} {\bibinfo {author} {\bibfnamefont {X.-P.}\ \bibnamefont
  {Li}}, \bibinfo {author} {\bibfnamefont {D.-S.}\ \bibnamefont {Ma}}, \bibinfo
  {author} {\bibfnamefont {C.-C.}\ \bibnamefont {Liu}}, \bibinfo {author}
  {\bibfnamefont {Z.-M.}\ \bibnamefont {Yu}},\ and\ \bibinfo {author}
  {\bibfnamefont {Y.}~\bibnamefont {Yao}},\ }\bibfield  {title} {\bibinfo
  {title} {From atomic semimetal to topological nontrivial insulator},\ }\href
  {https://doi.org/10.1103/PhysRevB.105.165135} {\bibfield  {journal} {\bibinfo
   {journal} {Phys. Rev. B}\ }\textbf {\bibinfo {volume} {105}},\ \bibinfo
  {pages} {165135} (\bibinfo {year} {2022})}\BibitemShut {NoStop}%
\bibitem [{\citenamefont {Ma}\ \emph {et~al.}(2023)\citenamefont {Ma},
  \citenamefont {Yu}, \citenamefont {Li}, \citenamefont {Zhou},\ and\
  \citenamefont {Wang}}]{OAI_cornermodes}%
  \BibitemOpen
  \bibfield  {author} {\bibinfo {author} {\bibfnamefont {D.-S.}\ \bibnamefont
  {Ma}}, \bibinfo {author} {\bibfnamefont {K.}~\bibnamefont {Yu}}, \bibinfo
  {author} {\bibfnamefont {X.-P.}\ \bibnamefont {Li}}, \bibinfo {author}
  {\bibfnamefont {X.}~\bibnamefont {Zhou}},\ and\ \bibinfo {author}
  {\bibfnamefont {R.}~\bibnamefont {Wang}},\ }\bibfield  {title} {\bibinfo
  {title} {Obstructed atomic insulators with robust corner modes},\ }\href
  {https://doi.org/10.1103/PhysRevB.108.L100101} {\bibfield  {journal}
  {\bibinfo  {journal} {Phys. Rev. B}\ }\textbf {\bibinfo {volume} {108}},\
  \bibinfo {pages} {L100101} (\bibinfo {year} {2023})}\BibitemShut {NoStop}%
\bibitem [{\citenamefont {Benalcazar}\ \emph {et~al.}(2019)\citenamefont
  {Benalcazar}, \citenamefont {Li},\ and\ \citenamefont
  {Hughes}}]{PRB_corner_charge}%
  \BibitemOpen
  \bibfield  {author} {\bibinfo {author} {\bibfnamefont {W.~A.}\ \bibnamefont
  {Benalcazar}}, \bibinfo {author} {\bibfnamefont {T.}~\bibnamefont {Li}},\
  and\ \bibinfo {author} {\bibfnamefont {T.~L.}\ \bibnamefont {Hughes}},\
  }\bibfield  {title} {\bibinfo {title} {Quantization of fractional corner
  charge in ${C}_{n}$-symmetric higher-order topological crystalline
  insulators},\ }\href {https://doi.org/10.1103/PhysRevB.99.245151} {\bibfield
  {journal} {\bibinfo  {journal} {Phys. Rev. B}\ }\textbf {\bibinfo {volume}
  {99}},\ \bibinfo {pages} {245151} (\bibinfo {year} {2019})}\BibitemShut
  {NoStop}%
\bibitem [{\citenamefont {Wang}\ \emph {et~al.}(2022)\citenamefont {Wang},
  \citenamefont {Jiang}, \citenamefont {Liu}, \citenamefont {Zhang},
  \citenamefont {Li}, \citenamefont {Liu}, \citenamefont {Sun}, \citenamefont
  {Weng},\ and\ \citenamefont {Chen}}]{OAI_MS2Z4}%
  \BibitemOpen
  \bibfield  {author} {\bibinfo {author} {\bibfnamefont {L.}~\bibnamefont
  {Wang}}, \bibinfo {author} {\bibfnamefont {Y.}~\bibnamefont {Jiang}},
  \bibinfo {author} {\bibfnamefont {J.}~\bibnamefont {Liu}}, \bibinfo {author}
  {\bibfnamefont {S.}~\bibnamefont {Zhang}}, \bibinfo {author} {\bibfnamefont
  {J.}~\bibnamefont {Li}}, \bibinfo {author} {\bibfnamefont {P.}~\bibnamefont
  {Liu}}, \bibinfo {author} {\bibfnamefont {Y.}~\bibnamefont {Sun}}, \bibinfo
  {author} {\bibfnamefont {H.}~\bibnamefont {Weng}},\ and\ \bibinfo {author}
  {\bibfnamefont {X.-Q.}\ \bibnamefont {Chen}},\ }\bibfield  {title} {\bibinfo
  {title} {Two-dimensional obstructed atomic insulators with fractional corner
  charge in the ${MA}_{2}{Z}_{4}$ family},\ }\href
  {https://doi.org/10.1103/PhysRevB.106.155144} {\bibfield  {journal} {\bibinfo
   {journal} {Phys. Rev. B}\ }\textbf {\bibinfo {volume} {106}},\ \bibinfo
  {pages} {155144} (\bibinfo {year} {2022})}\BibitemShut {NoStop}%
\bibitem [{\citenamefont {Sheng}\ \emph {et~al.}(2024)\citenamefont {Sheng},
  \citenamefont {Xie}, \citenamefont {Wu}, \citenamefont {Weng}, \citenamefont
  {Dai}, \citenamefont {Bernevig}, \citenamefont {Fang},\ and\ \citenamefont
  {Wang}}]{OAI_majorana}%
  \BibitemOpen
  \bibfield  {author} {\bibinfo {author} {\bibfnamefont {H.}~\bibnamefont
  {Sheng}}, \bibinfo {author} {\bibfnamefont {Y.}~\bibnamefont {Xie}}, \bibinfo
  {author} {\bibfnamefont {Q.}~\bibnamefont {Wu}}, \bibinfo {author}
  {\bibfnamefont {H.}~\bibnamefont {Weng}}, \bibinfo {author} {\bibfnamefont
  {X.}~\bibnamefont {Dai}}, \bibinfo {author} {\bibfnamefont {B.~A.}\
  \bibnamefont {Bernevig}}, \bibinfo {author} {\bibfnamefont {Z.}~\bibnamefont
  {Fang}},\ and\ \bibinfo {author} {\bibfnamefont {Z.}~\bibnamefont {Wang}},\
  }\bibfield  {title} {\bibinfo {title} {Majorana corner modes in
  unconventional monolayers of the ${1T}-{PtSe}_{2}$ family},\ }\href
  {https://doi.org/10.1103/PhysRevB.110.035151} {\bibfield  {journal} {\bibinfo
   {journal} {Phys. Rev. B}\ }\textbf {\bibinfo {volume} {110}},\ \bibinfo
  {pages} {035151} (\bibinfo {year} {2024})}\BibitemShut {NoStop}%
\bibitem [{\citenamefont {Yang}\ \emph {et~al.}(2024)\citenamefont {Yang},
  \citenamefont {Sheng}, \citenamefont {Guo}, \citenamefont {Zhang},
  \citenamefont {Wu}, \citenamefont {Weng}, \citenamefont {Fang},\ and\
  \citenamefont {Wang}}]{OAI_superconductivity}%
  \BibitemOpen
  \bibfield  {author} {\bibinfo {author} {\bibfnamefont {Z.}~\bibnamefont
  {Yang}}, \bibinfo {author} {\bibfnamefont {H.}~\bibnamefont {Sheng}},
  \bibinfo {author} {\bibfnamefont {Z.}~\bibnamefont {Guo}}, \bibinfo {author}
  {\bibfnamefont {R.}~\bibnamefont {Zhang}}, \bibinfo {author} {\bibfnamefont
  {Q.}~\bibnamefont {Wu}}, \bibinfo {author} {\bibfnamefont {H.}~\bibnamefont
  {Weng}}, \bibinfo {author} {\bibfnamefont {Z.}~\bibnamefont {Fang}},\ and\
  \bibinfo {author} {\bibfnamefont {Z.}~\bibnamefont {Wang}},\ }\bibfield
  {title} {\bibinfo {title} {Superconductivity in unconventional metals},\
  }\href {https://doi.org/10.1038/s41524-024-01210-z} {\bibfield  {journal}
  {\bibinfo  {journal} {Npj Comput. Mater.}\ }\textbf {\bibinfo {volume}
  {10}},\ \bibinfo {pages} {25} (\bibinfo {year} {2024})}\BibitemShut {NoStop}%
\bibitem [{\citenamefont {Eck}\ \emph {et~al.}(2022)\citenamefont {Eck},
  \citenamefont {Ortix}, \citenamefont {Consiglio}, \citenamefont {Erhardt},
  \citenamefont {Bauernfeind}, \citenamefont {Moser}, \citenamefont {Claessen},
  \citenamefont {Di~Sante},\ and\ \citenamefont {Sangiovanni}}]{OAI_QSH}%
  \BibitemOpen
  \bibfield  {author} {\bibinfo {author} {\bibfnamefont {P.}~\bibnamefont
  {Eck}}, \bibinfo {author} {\bibfnamefont {C.}~\bibnamefont {Ortix}}, \bibinfo
  {author} {\bibfnamefont {A.}~\bibnamefont {Consiglio}}, \bibinfo {author}
  {\bibfnamefont {J.}~\bibnamefont {Erhardt}}, \bibinfo {author} {\bibfnamefont
  {M.}~\bibnamefont {Bauernfeind}}, \bibinfo {author} {\bibfnamefont
  {S.}~\bibnamefont {Moser}}, \bibinfo {author} {\bibfnamefont
  {R.}~\bibnamefont {Claessen}}, \bibinfo {author} {\bibfnamefont
  {D.}~\bibnamefont {Di~Sante}},\ and\ \bibinfo {author} {\bibfnamefont
  {G.}~\bibnamefont {Sangiovanni}},\ }\bibfield  {title} {\bibinfo {title}
  {Real-space obstruction in quantum spin hall insulators},\ }\href
  {https://doi.org/10.1103/PhysRevB.106.195143} {\bibfield  {journal} {\bibinfo
   {journal} {Phys. Rev. B}\ }\textbf {\bibinfo {volume} {106}},\ \bibinfo
  {pages} {195143} (\bibinfo {year} {2022})}\BibitemShut {NoStop}%
\bibitem [{\citenamefont {Li}\ \emph {et~al.}()\citenamefont {Li},
  \citenamefont {Xu}, \citenamefont {Song}, \citenamefont {Yang}, \citenamefont
  {Zhang}, \citenamefont {Liu}, \citenamefont {Gupta}, \citenamefont {Sub},
  \citenamefont {Sun}, \citenamefont {Sessi}, \citenamefont {Parkin},
  \citenamefont {Bernevig},\ and\ \citenamefont {Felser}}]{OAI_active_sites}%
  \BibitemOpen
  \bibfield  {author} {\bibinfo {author} {\bibfnamefont {G.}~\bibnamefont
  {Li}}, \bibinfo {author} {\bibfnamefont {Y.}~\bibnamefont {Xu}}, \bibinfo
  {author} {\bibfnamefont {Z.}~\bibnamefont {Song}}, \bibinfo {author}
  {\bibfnamefont {Q.}~\bibnamefont {Yang}}, \bibinfo {author} {\bibfnamefont
  {Y.}~\bibnamefont {Zhang}}, \bibinfo {author} {\bibfnamefont
  {J.}~\bibnamefont {Liu}}, \bibinfo {author} {\bibfnamefont {U.}~\bibnamefont
  {Gupta}}, \bibinfo {author} {\bibfnamefont {V.}~\bibnamefont {Sub}}, \bibinfo
  {author} {\bibfnamefont {Y.}~\bibnamefont {Sun}}, \bibinfo {author}
  {\bibfnamefont {P.}~\bibnamefont {Sessi}}, \bibinfo {author} {\bibfnamefont
  {S.~S.~P.}\ \bibnamefont {Parkin}}, \bibinfo {author} {\bibfnamefont {B.~A.}\
  \bibnamefont {Bernevig}},\ and\ \bibinfo {author} {\bibfnamefont
  {C.}~\bibnamefont {Felser}},\ }\bibfield  {title} {\bibinfo {title}
  {Obstructed surface states as the descriptor for predicting catalytic active
  sites in inorganic crystalline materials},\ }\href
  {https://doi.org/https://doi.org/10.1002/adma.202201328} {\bibfield
  {journal} {\bibinfo  {journal} {Adv. Mater.}\ }\textbf {\bibinfo {volume}
  {34}},\ \bibinfo {pages} {2201328}}\BibitemShut {NoStop}%
\bibitem [{\citenamefont {Jiang}\ \emph {et~al.}(2023)\citenamefont {Jiang},
  \citenamefont {Qi}, \citenamefont {Weng},\ and\ \citenamefont
  {Hu}}]{OAI_mott}%
  \BibitemOpen
  \bibfield  {author} {\bibinfo {author} {\bibfnamefont {K.}~\bibnamefont
  {Jiang}}, \bibinfo {author} {\bibfnamefont {Z.}~\bibnamefont {Qi}}, \bibinfo
  {author} {\bibfnamefont {H.}~\bibnamefont {Weng}},\ and\ \bibinfo {author}
  {\bibfnamefont {J.}~\bibnamefont {Hu}},\ }\bibfield  {title} {\bibinfo
  {title} {Mottness in obstructed atomic insulators without mott transition},\
  }\href {https://doi.org/10.1103/PhysRevB.108.195102} {\bibfield  {journal}
  {\bibinfo  {journal} {Phys. Rev. B}\ }\textbf {\bibinfo {volume} {108}},\
  \bibinfo {pages} {195102} (\bibinfo {year} {2023})}\BibitemShut {NoStop}%
\bibitem [{\citenamefont {Novoselov}\ \emph {et~al.}(2016)\citenamefont
  {Novoselov}, \citenamefont {Mishchenko}, \citenamefont {Carvalho},\ and\
  \citenamefont {Neto}}]{2DvdW}%
  \BibitemOpen
  \bibfield  {author} {\bibinfo {author} {\bibfnamefont {K.~S.}\ \bibnamefont
  {Novoselov}}, \bibinfo {author} {\bibfnamefont {A.}~\bibnamefont
  {Mishchenko}}, \bibinfo {author} {\bibfnamefont {A.}~\bibnamefont
  {Carvalho}},\ and\ \bibinfo {author} {\bibfnamefont {A.~H.~C.}\ \bibnamefont
  {Neto}},\ }\bibfield  {title} {\bibinfo {title} {2{D} materials and van der
  waals heterostructures},\ }\href {https://doi.org/10.1126/science.aac9439}
  {\bibfield  {journal} {\bibinfo  {journal} {Science}\ }\textbf {\bibinfo
  {volume} {353}},\ \bibinfo {pages} {aac9439} (\bibinfo {year}
  {2016})}\BibitemShut {NoStop}%
\bibitem [{\citenamefont {Sheng}\ \emph {et~al.}(2021)\citenamefont {Sheng},
  \citenamefont {Hua}, \citenamefont {Cheng}, \citenamefont {Hu}, \citenamefont
  {Sun}, \citenamefont {Tao}, \citenamefont {Lu}, \citenamefont {Lu},
  \citenamefont {Zhong}, \citenamefont {Watanabe}, \citenamefont {Taniguchi},
  \citenamefont {Xia}, \citenamefont {Xu},\ and\ \citenamefont
  {Zheng}}]{Rasbha_As}%
  \BibitemOpen
  \bibfield  {author} {\bibinfo {author} {\bibfnamefont {F.}~\bibnamefont
  {Sheng}}, \bibinfo {author} {\bibfnamefont {C.}~\bibnamefont {Hua}}, \bibinfo
  {author} {\bibfnamefont {M.}~\bibnamefont {Cheng}}, \bibinfo {author}
  {\bibfnamefont {J.}~\bibnamefont {Hu}}, \bibinfo {author} {\bibfnamefont
  {X.}~\bibnamefont {Sun}}, \bibinfo {author} {\bibfnamefont {Q.}~\bibnamefont
  {Tao}}, \bibinfo {author} {\bibfnamefont {H.}~\bibnamefont {Lu}}, \bibinfo
  {author} {\bibfnamefont {Y.}~\bibnamefont {Lu}}, \bibinfo {author}
  {\bibfnamefont {M.}~\bibnamefont {Zhong}}, \bibinfo {author} {\bibfnamefont
  {K.}~\bibnamefont {Watanabe}}, \bibinfo {author} {\bibfnamefont
  {T.}~\bibnamefont {Taniguchi}}, \bibinfo {author} {\bibfnamefont
  {Q.}~\bibnamefont {Xia}}, \bibinfo {author} {\bibfnamefont {Z.-A.}\
  \bibnamefont {Xu}},\ and\ \bibinfo {author} {\bibfnamefont {Y.}~\bibnamefont
  {Zheng}},\ }\bibfield  {title} {\bibinfo {title} {Rashba valleys and quantum
  hall states in few-layer black arsenic},\ }\href
  {https://doi.org/10.1038/s41586-021-03449-8} {\bibfield  {journal} {\bibinfo
  {journal} {Nature}\ }\textbf {\bibinfo {volume} {593}},\ \bibinfo {pages}
  {56} (\bibinfo {year} {2021})}\BibitemShut {NoStop}%
\bibitem [{\citenamefont {Carvalho}\ \emph {et~al.}(2016)\citenamefont
  {Carvalho}, \citenamefont {Wang}, \citenamefont {Zhu}, \citenamefont {Rodin},
  \citenamefont {Su},\ and\ \citenamefont {Castro~Neto}}]{phosphorene}%
  \BibitemOpen
  \bibfield  {author} {\bibinfo {author} {\bibfnamefont {A.}~\bibnamefont
  {Carvalho}}, \bibinfo {author} {\bibfnamefont {M.}~\bibnamefont {Wang}},
  \bibinfo {author} {\bibfnamefont {X.}~\bibnamefont {Zhu}}, \bibinfo {author}
  {\bibfnamefont {A.~S.}\ \bibnamefont {Rodin}}, \bibinfo {author}
  {\bibfnamefont {H.}~\bibnamefont {Su}},\ and\ \bibinfo {author}
  {\bibfnamefont {A.~H.}\ \bibnamefont {Castro~Neto}},\ }\bibfield  {title}
  {\bibinfo {title} {Phosphorene: from theory to applications},\ }\href
  {https://doi.org/10.1038/natrevmats.2016.61} {\bibfield  {journal} {\bibinfo
  {journal} {Nat Rev Mater}\ }\textbf {\bibinfo {volume} {1}},\ \bibinfo
  {pages} {16061} (\bibinfo {year} {2016})}\BibitemShut {NoStop}%
\bibitem [{\citenamefont {Xia}\ \emph {et~al.}(2019)\citenamefont {Xia},
  \citenamefont {Wang}, \citenamefont {Hwang}, \citenamefont {Neto},\ and\
  \citenamefont {Yang}}]{P_isoelectron}%
  \BibitemOpen
  \bibfield  {author} {\bibinfo {author} {\bibfnamefont {F.}~\bibnamefont
  {Xia}}, \bibinfo {author} {\bibfnamefont {H.}~\bibnamefont {Wang}}, \bibinfo
  {author} {\bibfnamefont {J.~C.~M.}\ \bibnamefont {Hwang}}, \bibinfo {author}
  {\bibfnamefont {A.~H.~C.}\ \bibnamefont {Neto}},\ and\ \bibinfo {author}
  {\bibfnamefont {L.}~\bibnamefont {Yang}},\ }\bibfield  {title} {\bibinfo
  {title} {Black phosphorus and its isoelectronic materials},\ }\href
  {https://doi.org/10.1038/s42254-019-0043-5} {\bibfield  {journal} {\bibinfo
  {journal} {Nat. Rev. Phys.}\ }\textbf {\bibinfo {volume} {1}},\ \bibinfo
  {pages} {306} (\bibinfo {year} {2019})}\BibitemShut {NoStop}%
\bibitem [{\citenamefont {Kim}\ \emph {et~al.}(2015)\citenamefont {Kim},
  \citenamefont {Baik}, \citenamefont {Ryu}, \citenamefont {Sohn},
  \citenamefont {Park}, \citenamefont {Park}, \citenamefont {Denlinger},
  \citenamefont {Yi}, \citenamefont {Choi},\ and\ \citenamefont
  {Kim}}]{phos_science}%
  \BibitemOpen
  \bibfield  {author} {\bibinfo {author} {\bibfnamefont {J.}~\bibnamefont
  {Kim}}, \bibinfo {author} {\bibfnamefont {S.~S.}\ \bibnamefont {Baik}},
  \bibinfo {author} {\bibfnamefont {S.~H.}\ \bibnamefont {Ryu}}, \bibinfo
  {author} {\bibfnamefont {Y.}~\bibnamefont {Sohn}}, \bibinfo {author}
  {\bibfnamefont {S.}~\bibnamefont {Park}}, \bibinfo {author} {\bibfnamefont
  {B.-G.}\ \bibnamefont {Park}}, \bibinfo {author} {\bibfnamefont
  {J.}~\bibnamefont {Denlinger}}, \bibinfo {author} {\bibfnamefont
  {Y.}~\bibnamefont {Yi}}, \bibinfo {author} {\bibfnamefont {H.~J.}\
  \bibnamefont {Choi}},\ and\ \bibinfo {author} {\bibfnamefont {K.~S.}\
  \bibnamefont {Kim}},\ }\bibfield  {title} {\bibinfo {title} {Observation of
  tunable band gap and anisotropic dirac semimetal state in black phosphorus},\
  }\href {https://doi.org/10.1126/science.aaa6486} {\bibfield  {journal}
  {\bibinfo  {journal} {Science}\ }\textbf {\bibinfo {volume} {349}},\ \bibinfo
  {pages} {723} (\bibinfo {year} {2015})}\BibitemShut {NoStop}%
\bibitem [{\citenamefont {Lu}\ \emph {et~al.}(2021)\citenamefont {Lu},
  \citenamefont {Chen}, \citenamefont {Snyder}, \citenamefont {Cook},
  \citenamefont {Nguyen}, \citenamefont {Reddy}, \citenamefont {Chang},
  \citenamefont {Yang},\ and\ \citenamefont {Bian}}]{PRB2021_Sb}%
  \BibitemOpen
  \bibfield  {author} {\bibinfo {author} {\bibfnamefont {Q.}~\bibnamefont
  {Lu}}, \bibinfo {author} {\bibfnamefont {K.~Y.}\ \bibnamefont {Chen}},
  \bibinfo {author} {\bibfnamefont {M.}~\bibnamefont {Snyder}}, \bibinfo
  {author} {\bibfnamefont {J.}~\bibnamefont {Cook}}, \bibinfo {author}
  {\bibfnamefont {D.~T.}\ \bibnamefont {Nguyen}}, \bibinfo {author}
  {\bibfnamefont {P.~V.~S.}\ \bibnamefont {Reddy}}, \bibinfo {author}
  {\bibfnamefont {T.-R.}\ \bibnamefont {Chang}}, \bibinfo {author}
  {\bibfnamefont {S.~A.}\ \bibnamefont {Yang}},\ and\ \bibinfo {author}
  {\bibfnamefont {G.}~\bibnamefont {Bian}},\ }\bibfield  {title} {\bibinfo
  {title} {Observation of symmetry-protected dirac states in nonsymmorphic
  $\ensuremath{\alpha}$-antimonene},\ }\href
  {https://doi.org/10.1103/PhysRevB.104.L201105} {\bibfield  {journal}
  {\bibinfo  {journal} {Phys. Rev. B}\ }\textbf {\bibinfo {volume} {104}},\
  \bibinfo {pages} {L201105} (\bibinfo {year} {2021})}\BibitemShut {NoStop}%
\bibitem [{\citenamefont {Kowalczyk}\ \emph {et~al.}(2020)\citenamefont
  {Kowalczyk}, \citenamefont {Brown}, \citenamefont {Maerkl}, \citenamefont
  {Lu}, \citenamefont {Chiu}, \citenamefont {Liu}, \citenamefont {Yang},
  \citenamefont {Wang}, \citenamefont {Zasada}, \citenamefont {Genuzio},
  \citenamefont {Menteş}, \citenamefont {Locatelli}, \citenamefont {Chiang},\
  and\ \citenamefont {Bian}}]{alpha_Bi}%
  \BibitemOpen
  \bibfield  {author} {\bibinfo {author} {\bibfnamefont {P.~J.}\ \bibnamefont
  {Kowalczyk}}, \bibinfo {author} {\bibfnamefont {S.~A.}\ \bibnamefont
  {Brown}}, \bibinfo {author} {\bibfnamefont {T.}~\bibnamefont {Maerkl}},
  \bibinfo {author} {\bibfnamefont {Q.}~\bibnamefont {Lu}}, \bibinfo {author}
  {\bibfnamefont {C.-K.}\ \bibnamefont {Chiu}}, \bibinfo {author}
  {\bibfnamefont {Y.}~\bibnamefont {Liu}}, \bibinfo {author} {\bibfnamefont
  {S.~A.}\ \bibnamefont {Yang}}, \bibinfo {author} {\bibfnamefont
  {X.}~\bibnamefont {Wang}}, \bibinfo {author} {\bibfnamefont {I.}~\bibnamefont
  {Zasada}}, \bibinfo {author} {\bibfnamefont {F.}~\bibnamefont {Genuzio}},
  \bibinfo {author} {\bibfnamefont {T.~O.}\ \bibnamefont {Menteş}}, \bibinfo
  {author} {\bibfnamefont {A.}~\bibnamefont {Locatelli}}, \bibinfo {author}
  {\bibfnamefont {T.-C.}\ \bibnamefont {Chiang}},\ and\ \bibinfo {author}
  {\bibfnamefont {G.}~\bibnamefont {Bian}},\ }\bibfield  {title} {\bibinfo
  {title} {Realization of symmetry-enforced two-dimensional dirac fermions in
  nonsymmorphic $\alpha$-bismuthene},\ }\href
  {https://doi.org/10.1021/acsnano.9b08136} {\bibfield  {journal} {\bibinfo
  {journal} {ACS Nano}\ }\textbf {\bibinfo {volume} {14}},\ \bibinfo {pages}
  {1888} (\bibinfo {year} {2020})}\BibitemShut {NoStop}%
\bibitem [{\citenamefont {Islam}\ \emph {et~al.}(2021)\citenamefont {Islam},
  \citenamefont {Ghosh}, \citenamefont {Autieri}, \citenamefont {Chowdhury},
  \citenamefont {Bansil}, \citenamefont {Agarwal},\ and\ \citenamefont
  {Singh}}]{Rajibul_MSi2Z4}%
  \BibitemOpen
  \bibfield  {author} {\bibinfo {author} {\bibfnamefont {R.}~\bibnamefont
  {Islam}}, \bibinfo {author} {\bibfnamefont {B.}~\bibnamefont {Ghosh}},
  \bibinfo {author} {\bibfnamefont {C.}~\bibnamefont {Autieri}}, \bibinfo
  {author} {\bibfnamefont {S.}~\bibnamefont {Chowdhury}}, \bibinfo {author}
  {\bibfnamefont {A.}~\bibnamefont {Bansil}}, \bibinfo {author} {\bibfnamefont
  {A.}~\bibnamefont {Agarwal}},\ and\ \bibinfo {author} {\bibfnamefont
  {B.}~\bibnamefont {Singh}},\ }\bibfield  {title} {\bibinfo {title} {Tunable
  spin polarization and electronic structure of bottom-up synthesized
  {MoSi}$_{2}${N}$_{4}$ materials},\ }\href
  {https://doi.org/10.1103/PhysRevB.104.L201112} {\bibfield  {journal}
  {\bibinfo  {journal} {Phys. Rev. B}\ }\textbf {\bibinfo {volume} {104}},\
  \bibinfo {pages} {L201112} (\bibinfo {year} {2021})}\BibitemShut {NoStop}%
\bibitem [{\citenamefont {Lu}\ \emph {et~al.}(2022)\citenamefont {Lu},
  \citenamefont {Cook}, \citenamefont {Zhang}, \citenamefont {Chen},
  \citenamefont {Snyder}, \citenamefont {Nguyen}, \citenamefont {Reddy},
  \citenamefont {Qin}, \citenamefont {Zhan}, \citenamefont {Zhao},
  \citenamefont {Kowalczyk}, \citenamefont {Brown}, \citenamefont {Chiang},
  \citenamefont {Yang}, \citenamefont {Chang},\ and\ \citenamefont
  {Bian}}]{Unpinned_Sb}%
  \BibitemOpen
  \bibfield  {author} {\bibinfo {author} {\bibfnamefont {Q.}~\bibnamefont
  {Lu}}, \bibinfo {author} {\bibfnamefont {J.}~\bibnamefont {Cook}}, \bibinfo
  {author} {\bibfnamefont {X.}~\bibnamefont {Zhang}}, \bibinfo {author}
  {\bibfnamefont {K.~Y.}\ \bibnamefont {Chen}}, \bibinfo {author}
  {\bibfnamefont {M.}~\bibnamefont {Snyder}}, \bibinfo {author} {\bibfnamefont
  {D.~T.}\ \bibnamefont {Nguyen}}, \bibinfo {author} {\bibfnamefont {P.~V.~S.}\
  \bibnamefont {Reddy}}, \bibinfo {author} {\bibfnamefont {B.}~\bibnamefont
  {Qin}}, \bibinfo {author} {\bibfnamefont {S.}~\bibnamefont {Zhan}}, \bibinfo
  {author} {\bibfnamefont {L.-D.}\ \bibnamefont {Zhao}}, \bibinfo {author}
  {\bibfnamefont {P.~J.}\ \bibnamefont {Kowalczyk}}, \bibinfo {author}
  {\bibfnamefont {S.~A.}\ \bibnamefont {Brown}}, \bibinfo {author}
  {\bibfnamefont {T.-C.}\ \bibnamefont {Chiang}}, \bibinfo {author}
  {\bibfnamefont {S.~A.}\ \bibnamefont {Yang}}, \bibinfo {author}
  {\bibfnamefont {T.-R.}\ \bibnamefont {Chang}},\ and\ \bibinfo {author}
  {\bibfnamefont {G.}~\bibnamefont {Bian}},\ }\bibfield  {title} {\bibinfo
  {title} {Realization of unpinned two-dimensional dirac states in antimony
  atomic layers},\ }\href {https://doi.org/10.1038/s41467-022-32327-8}
  {\bibfield  {journal} {\bibinfo  {journal} {Nat. Commun.}\ }\textbf {\bibinfo
  {volume} {13}},\ \bibinfo {pages} {4603} (\bibinfo {year}
  {2022})}\BibitemShut {NoStop}%
\bibitem [{\citenamefont {Ghosh}\ \emph {et~al.}(2016)\citenamefont {Ghosh},
  \citenamefont {Singh}, \citenamefont {Prasad},\ and\ \citenamefont
  {Agarwal}}]{Ghosh2016}%
  \BibitemOpen
  \bibfield  {author} {\bibinfo {author} {\bibfnamefont {B.}~\bibnamefont
  {Ghosh}}, \bibinfo {author} {\bibfnamefont {B.}~\bibnamefont {Singh}},
  \bibinfo {author} {\bibfnamefont {R.}~\bibnamefont {Prasad}},\ and\ \bibinfo
  {author} {\bibfnamefont {A.}~\bibnamefont {Agarwal}},\ }\bibfield  {title}
  {\bibinfo {title} {Electric-field tunable dirac semimetal state in
  phosphorene thin films},\ }\href {https://doi.org/10.1103/PhysRevB.94.205426}
  {\bibfield  {journal} {\bibinfo  {journal} {Phys. Rev. B}\ }\textbf {\bibinfo
  {volume} {94}},\ \bibinfo {pages} {205426} (\bibinfo {year}
  {2016})}\BibitemShut {NoStop}%
\bibitem [{\citenamefont {Baik}\ \emph {et~al.}(2015)\citenamefont {Baik},
  \citenamefont {Kim}, \citenamefont {Yi},\ and\ \citenamefont
  {Choi}}]{phos_doping}%
  \BibitemOpen
  \bibfield  {author} {\bibinfo {author} {\bibfnamefont {S.~S.}\ \bibnamefont
  {Baik}}, \bibinfo {author} {\bibfnamefont {K.~S.}\ \bibnamefont {Kim}},
  \bibinfo {author} {\bibfnamefont {Y.}~\bibnamefont {Yi}},\ and\ \bibinfo
  {author} {\bibfnamefont {H.~J.}\ \bibnamefont {Choi}},\ }\bibfield  {title}
  {\bibinfo {title} {Emergence of two-dimensional massless dirac fermions,
  chiral pseudospins, and berry’s phase in potassium doped few-layer black
  phosphorus},\ }\href {https://doi.org/10.1021/acs.nanolett.5b04106}
  {\bibfield  {journal} {\bibinfo  {journal} {Nano Lett.}\ }\textbf {\bibinfo
  {volume} {15}},\ \bibinfo {pages} {7788} (\bibinfo {year}
  {2015})}\BibitemShut {NoStop}%
\bibitem [{\citenamefont {Liu}\ \emph {et~al.}(2015)\citenamefont {Liu},
  \citenamefont {Zhang}, \citenamefont {Abdalla}, \citenamefont {Fazzio},\ and\
  \citenamefont {Zunger}}]{phos_Efield}%
  \BibitemOpen
  \bibfield  {author} {\bibinfo {author} {\bibfnamefont {Q.}~\bibnamefont
  {Liu}}, \bibinfo {author} {\bibfnamefont {X.}~\bibnamefont {Zhang}}, \bibinfo
  {author} {\bibfnamefont {L.~B.}\ \bibnamefont {Abdalla}}, \bibinfo {author}
  {\bibfnamefont {A.}~\bibnamefont {Fazzio}},\ and\ \bibinfo {author}
  {\bibfnamefont {A.}~\bibnamefont {Zunger}},\ }\bibfield  {title} {\bibinfo
  {title} {Switching a normal insulator into a topological insulator via
  electric field with application to phosphorene},\ }\href
  {https://doi.org/10.1021/nl5043769} {\bibfield  {journal} {\bibinfo
  {journal} {Nano Lett.}\ }\textbf {\bibinfo {volume} {15}},\ \bibinfo {pages}
  {1222} (\bibinfo {year} {2015})}\BibitemShut {NoStop}%
\bibitem [{\citenamefont {Xiang}\ \emph {et~al.}(2015)\citenamefont {Xiang},
  \citenamefont {Ye}, \citenamefont {Shang}, \citenamefont {Lei}, \citenamefont
  {Wang}, \citenamefont {Yang}, \citenamefont {Liu}, \citenamefont {Meng},
  \citenamefont {Luo}, \citenamefont {Zou}, \citenamefont {Sun}, \citenamefont
  {Zhang},\ and\ \citenamefont {Chen}}]{phos_pressure}%
  \BibitemOpen
  \bibfield  {author} {\bibinfo {author} {\bibfnamefont {Z.~J.}\ \bibnamefont
  {Xiang}}, \bibinfo {author} {\bibfnamefont {G.~J.}\ \bibnamefont {Ye}},
  \bibinfo {author} {\bibfnamefont {C.}~\bibnamefont {Shang}}, \bibinfo
  {author} {\bibfnamefont {B.}~\bibnamefont {Lei}}, \bibinfo {author}
  {\bibfnamefont {N.~Z.}\ \bibnamefont {Wang}}, \bibinfo {author}
  {\bibfnamefont {K.~S.}\ \bibnamefont {Yang}}, \bibinfo {author}
  {\bibfnamefont {D.~Y.}\ \bibnamefont {Liu}}, \bibinfo {author} {\bibfnamefont
  {F.~B.}\ \bibnamefont {Meng}}, \bibinfo {author} {\bibfnamefont {X.~G.}\
  \bibnamefont {Luo}}, \bibinfo {author} {\bibfnamefont {L.~J.}\ \bibnamefont
  {Zou}}, \bibinfo {author} {\bibfnamefont {Z.}~\bibnamefont {Sun}}, \bibinfo
  {author} {\bibfnamefont {Y.}~\bibnamefont {Zhang}},\ and\ \bibinfo {author}
  {\bibfnamefont {X.~H.}\ \bibnamefont {Chen}},\ }\bibfield  {title} {\bibinfo
  {title} {Pressure-induced electronic transition in black phosphorus},\ }\href
  {https://doi.org/10.1103/PhysRevLett.115.186403} {\bibfield  {journal}
  {\bibinfo  {journal} {Phys. Rev. Lett.}\ }\textbf {\bibinfo {volume} {115}},\
  \bibinfo {pages} {186403} (\bibinfo {year} {2015})}\BibitemShut {NoStop}%
\bibitem [{\citenamefont {Lu}\ \emph {et~al.}(2016)\citenamefont {Lu},
  \citenamefont {Zhou}, \citenamefont {Chang}, \citenamefont {Guan},
  \citenamefont {Chen}, \citenamefont {Jiang}, \citenamefont {Jiang},
  \citenamefont {sen Wang}, \citenamefont {Yang}, \citenamefont {Feng},
  \citenamefont {Kawazoe},\ and\ \citenamefont {Lin}}]{Unpinned_Dirac}%
  \BibitemOpen
  \bibfield  {author} {\bibinfo {author} {\bibfnamefont {Y.}~\bibnamefont
  {Lu}}, \bibinfo {author} {\bibfnamefont {D.}~\bibnamefont {Zhou}}, \bibinfo
  {author} {\bibfnamefont {G.}~\bibnamefont {Chang}}, \bibinfo {author}
  {\bibfnamefont {S.}~\bibnamefont {Guan}}, \bibinfo {author} {\bibfnamefont
  {W.}~\bibnamefont {Chen}}, \bibinfo {author} {\bibfnamefont {Y.}~\bibnamefont
  {Jiang}}, \bibinfo {author} {\bibfnamefont {J.}~\bibnamefont {Jiang}},
  \bibinfo {author} {\bibfnamefont {X.}~\bibnamefont {sen Wang}}, \bibinfo
  {author} {\bibfnamefont {S.~A.}\ \bibnamefont {Yang}}, \bibinfo {author}
  {\bibfnamefont {Y.~P.}\ \bibnamefont {Feng}}, \bibinfo {author}
  {\bibfnamefont {Y.}~\bibnamefont {Kawazoe}},\ and\ \bibinfo {author}
  {\bibfnamefont {H.}~\bibnamefont {Lin}},\ }\bibfield  {title} {\bibinfo
  {title} {Multiple unpinned dirac points in group-{V}a single-layers with
  phosphorene structure},\ }\href
  {https://doi.org/10.1038/npjcompumats.2016.11} {\bibfield  {journal}
  {\bibinfo  {journal} {Npj Comput. Mater.}\ }\textbf {\bibinfo {volume} {2}},\
  \bibinfo {pages} {16011} (\bibinfo {year} {2016})}\BibitemShut {NoStop}%
\bibitem [{\citenamefont {Wang}\ \emph {et~al.}(2024)\citenamefont {Wang},
  \citenamefont {Zhou}, \citenamefont {Hung}, \citenamefont {Lin},
  \citenamefont {Lin},\ and\ \citenamefont {Bansil}}]{HSCI_BaoKai}%
  \BibitemOpen
  \bibfield  {author} {\bibinfo {author} {\bibfnamefont {B.}~\bibnamefont
  {Wang}}, \bibinfo {author} {\bibfnamefont {X.}~\bibnamefont {Zhou}}, \bibinfo
  {author} {\bibfnamefont {Y.-C.}\ \bibnamefont {Hung}}, \bibinfo {author}
  {\bibfnamefont {Y.-C.}\ \bibnamefont {Lin}}, \bibinfo {author} {\bibfnamefont
  {H.}~\bibnamefont {Lin}},\ and\ \bibinfo {author} {\bibfnamefont
  {A.}~\bibnamefont {Bansil}},\ }\bibfield  {title} {\bibinfo {title} {High
  spin-chern-number insulator in $\alpha$-antimonene with a hidden topological
  phase},\ }\href {https://doi.org/10.1088/2053-1583/ad3136} {\bibfield
  {journal} {\bibinfo  {journal} {2D Mater.}\ }\textbf {\bibinfo {volume}
  {11}},\ \bibinfo {pages} {025033} (\bibinfo {year} {2024})}\BibitemShut
  {NoStop}%
\bibitem [{\citenamefont {Guo}\ \emph {et~al.}(2022)\citenamefont {Guo},
  \citenamefont {Hu}, \citenamefont {Putzke}, \citenamefont {Diaz},
  \citenamefont {Huang}, \citenamefont {Manna}, \citenamefont {Fan},
  \citenamefont {Shekhar}, \citenamefont {Sun}, \citenamefont {Felser},
  \citenamefont {Liu}, \citenamefont {Bernevig},\ and\ \citenamefont
  {Moll}}]{QuasiSymm}%
  \BibitemOpen
  \bibfield  {author} {\bibinfo {author} {\bibfnamefont {C.}~\bibnamefont
  {Guo}}, \bibinfo {author} {\bibfnamefont {L.}~\bibnamefont {Hu}}, \bibinfo
  {author} {\bibfnamefont {C.}~\bibnamefont {Putzke}}, \bibinfo {author}
  {\bibfnamefont {J.}~\bibnamefont {Diaz}}, \bibinfo {author} {\bibfnamefont
  {X.}~\bibnamefont {Huang}}, \bibinfo {author} {\bibfnamefont
  {K.}~\bibnamefont {Manna}}, \bibinfo {author} {\bibfnamefont {F.-R.}\
  \bibnamefont {Fan}}, \bibinfo {author} {\bibfnamefont {C.}~\bibnamefont
  {Shekhar}}, \bibinfo {author} {\bibfnamefont {Y.}~\bibnamefont {Sun}},
  \bibinfo {author} {\bibfnamefont {C.}~\bibnamefont {Felser}}, \bibinfo
  {author} {\bibfnamefont {C.}~\bibnamefont {Liu}}, \bibinfo {author}
  {\bibfnamefont {B.~A.}\ \bibnamefont {Bernevig}},\ and\ \bibinfo {author}
  {\bibfnamefont {P.~J.~W.}\ \bibnamefont {Moll}},\ }\bibfield  {title}
  {\bibinfo {title} {Quasi-symmetry-protected topology in a semi-metal},\
  }\href {https://doi.org/10.1038/s41567-022-01604-0} {\bibfield  {journal}
  {\bibinfo  {journal} {Nat. Phys.}\ }\textbf {\bibinfo {volume} {18}},\
  \bibinfo {pages} {813–818} (\bibinfo {year} {2022})}\BibitemShut {NoStop}%
\bibitem [{\citenamefont {Li}\ \emph {et~al.}(2024)\citenamefont {Li},
  \citenamefont {Zhang}, \citenamefont {Liu},\ and\ \citenamefont
  {Liu}}]{QuasiSymm_GT}%
  \BibitemOpen
  \bibfield  {author} {\bibinfo {author} {\bibfnamefont {J.}~\bibnamefont
  {Li}}, \bibinfo {author} {\bibfnamefont {A.}~\bibnamefont {Zhang}}, \bibinfo
  {author} {\bibfnamefont {Y.}~\bibnamefont {Liu}},\ and\ \bibinfo {author}
  {\bibfnamefont {Q.}~\bibnamefont {Liu}},\ }\bibfield  {title} {\bibinfo
  {title} {Group theory on quasisymmetry and protected near degeneracy},\
  }\href {https://doi.org/10.1103/PhysRevLett.133.026402} {\bibfield  {journal}
  {\bibinfo  {journal} {Phys. Rev. Lett.}\ }\textbf {\bibinfo {volume} {133}},\
  \bibinfo {pages} {026402} (\bibinfo {year} {2024})}\BibitemShut {NoStop}%
\bibitem [{\citenamefont {Liu}\ \emph {et~al.}(2024{\natexlab{b}})\citenamefont
  {Liu}, \citenamefont {Liu}, \citenamefont {Li}, \citenamefont {Wu},\ and\
  \citenamefont {Liu}}]{QuasiSpin}%
  \BibitemOpen
  \bibfield  {author} {\bibinfo {author} {\bibfnamefont {L.}~\bibnamefont
  {Liu}}, \bibinfo {author} {\bibfnamefont {Y.}~\bibnamefont {Liu}}, \bibinfo
  {author} {\bibfnamefont {J.}~\bibnamefont {Li}}, \bibinfo {author}
  {\bibfnamefont {H.}~\bibnamefont {Wu}},\ and\ \bibinfo {author}
  {\bibfnamefont {Q.}~\bibnamefont {Liu}},\ }\bibfield  {title} {\bibinfo
  {title} {Quantum spin hall effect protected by spin ${U}$(1)
  quasi-symmetry},\ }\href {https://doi.org/10.48550/arXiv.2402.13974} {\
  (\bibinfo {year} {2024}{\natexlab{b}})},\ \Eprint
  {https://arxiv.org/abs/2402.13974} {arXiv:2402.13974} \BibitemShut {NoStop}%
\bibitem [{\citenamefont {Liu}\ \emph {et~al.}(2024{\natexlab{c}})\citenamefont
  {Liu}, \citenamefont {Liu}, \citenamefont {Li}, \citenamefont {Wu},\ and\
  \citenamefont {Liu}}]{Quasi_ESCI}%
  \BibitemOpen
  \bibfield  {author} {\bibinfo {author} {\bibfnamefont {L.}~\bibnamefont
  {Liu}}, \bibinfo {author} {\bibfnamefont {Y.}~\bibnamefont {Liu}}, \bibinfo
  {author} {\bibfnamefont {J.}~\bibnamefont {Li}}, \bibinfo {author}
  {\bibfnamefont {H.}~\bibnamefont {Wu}},\ and\ \bibinfo {author}
  {\bibfnamefont {Q.}~\bibnamefont {Liu}},\ }\bibfield  {title} {\bibinfo
  {title} {Orbital doublet driven even-spin chern insulators},\ }\href
  {https://doi.org/10.1103/PhysRevB.110.035161} {\bibfield  {journal} {\bibinfo
   {journal} {Phys. Rev. B}\ }\textbf {\bibinfo {volume} {110}},\ \bibinfo
  {pages} {035161} (\bibinfo {year} {2024}{\natexlab{c}})}\BibitemShut
  {NoStop}%
\bibitem [{sup()}]{supp}%
  \BibitemOpen
  \href@noop {} {}\bibinfo {note} {See Supplemental Materials for computational
  and crystal structure details, methodology for band representations,
  doping-dependent electronic structure, and spin-texture of edge states. The
  Supplementary Materials also contains
  Refs.~\cite{kohan_dft,VASP,perdew1996generalized,Grimme2010,mostofi2008wannier90,wu2018wanniertools,Greenwanniertools,Unpinned_Dirac,QuasiSpin,irvsp,bilbao_1,bilbao_2,Bilbao_online}}\BibitemShut
  {NoStop}%
\bibitem [{\citenamefont {Sun}\ \emph {et~al.}(2016)\citenamefont {Sun},
  \citenamefont {Zhang}, \citenamefont {Felser},\ and\ \citenamefont
  {Yan}}]{shc_TaAs}%
  \BibitemOpen
  \bibfield  {author} {\bibinfo {author} {\bibfnamefont {Y.}~\bibnamefont
  {Sun}}, \bibinfo {author} {\bibfnamefont {Y.}~\bibnamefont {Zhang}}, \bibinfo
  {author} {\bibfnamefont {C.}~\bibnamefont {Felser}},\ and\ \bibinfo {author}
  {\bibfnamefont {B.}~\bibnamefont {Yan}},\ }\bibfield  {title} {\bibinfo
  {title} {Strong intrinsic spin hall effect in the {TaAs} family of weyl
  semimetals},\ }\href {https://doi.org/10.1103/PhysRevLett.117.146403}
  {\bibfield  {journal} {\bibinfo  {journal} {Phys. Rev. Lett.}\ }\textbf
  {\bibinfo {volume} {117}},\ \bibinfo {pages} {146403} (\bibinfo {year}
  {2016})}\BibitemShut {NoStop}%
\bibitem [{\citenamefont {Qiao}\ \emph {et~al.}(2018)\citenamefont {Qiao},
  \citenamefont {Zhou}, \citenamefont {Yuan},\ and\ \citenamefont
  {Zhao}}]{SHC_w90}%
  \BibitemOpen
  \bibfield  {author} {\bibinfo {author} {\bibfnamefont {J.}~\bibnamefont
  {Qiao}}, \bibinfo {author} {\bibfnamefont {J.}~\bibnamefont {Zhou}}, \bibinfo
  {author} {\bibfnamefont {Z.}~\bibnamefont {Yuan}},\ and\ \bibinfo {author}
  {\bibfnamefont {W.}~\bibnamefont {Zhao}},\ }\bibfield  {title} {\bibinfo
  {title} {Calculation of intrinsic spin hall conductivity by wannier
  interpolation},\ }\href {https://doi.org/10.1103/PhysRevB.98.214402}
  {\bibfield  {journal} {\bibinfo  {journal} {Phys. Rev. B}\ }\textbf {\bibinfo
  {volume} {98}},\ \bibinfo {pages} {214402} (\bibinfo {year}
  {2018})}\BibitemShut {NoStop}%
\bibitem [{\citenamefont {Kane}\ and\ \citenamefont
  {Mele}(2005)}]{Kane_mele_QSH}%
  \BibitemOpen
  \bibfield  {author} {\bibinfo {author} {\bibfnamefont {C.~L.}\ \bibnamefont
  {Kane}}\ and\ \bibinfo {author} {\bibfnamefont {E.~J.}\ \bibnamefont
  {Mele}},\ }\bibfield  {title} {\bibinfo {title} {${Z}_{2}$ topological order
  and the quantum spin hall effect},\ }\href
  {https://doi.org/10.1103/PhysRevLett.95.146802} {\bibfield  {journal}
  {\bibinfo  {journal} {Phys. Rev. Lett.}\ }\textbf {\bibinfo {volume} {95}},\
  \bibinfo {pages} {146802} (\bibinfo {year} {2005})}\BibitemShut {NoStop}%
\bibitem [{\citenamefont {Sheng}\ \emph {et~al.}(2006)\citenamefont {Sheng},
  \citenamefont {Weng}, \citenamefont {Sheng},\ and\ \citenamefont
  {Haldane}}]{PRL2006}%
  \BibitemOpen
  \bibfield  {author} {\bibinfo {author} {\bibfnamefont {D.~N.}\ \bibnamefont
  {Sheng}}, \bibinfo {author} {\bibfnamefont {Z.~Y.}\ \bibnamefont {Weng}},
  \bibinfo {author} {\bibfnamefont {L.}~\bibnamefont {Sheng}},\ and\ \bibinfo
  {author} {\bibfnamefont {F.~D.~M.}\ \bibnamefont {Haldane}},\ }\bibfield
  {title} {\bibinfo {title} {Quantum spin-hall effect and topologically
  invariant chern numbers},\ }\href
  {https://doi.org/10.1103/PhysRevLett.97.036808} {\bibfield  {journal}
  {\bibinfo  {journal} {Phys. Rev. Lett.}\ }\textbf {\bibinfo {volume} {97}},\
  \bibinfo {pages} {036808} (\bibinfo {year} {2006})}\BibitemShut {NoStop}%
\bibitem [{\citenamefont {Prodan}(2009)}]{Pordan_2009}%
  \BibitemOpen
  \bibfield  {author} {\bibinfo {author} {\bibfnamefont {E.}~\bibnamefont
  {Prodan}},\ }\bibfield  {title} {\bibinfo {title} {Robustness of the
  spin-chern number},\ }\href {https://doi.org/10.1103/PhysRevB.80.125327}
  {\bibfield  {journal} {\bibinfo  {journal} {Phys. Rev. B}\ }\textbf {\bibinfo
  {volume} {80}},\ \bibinfo {pages} {125327} (\bibinfo {year}
  {2009})}\BibitemShut {NoStop}%
\bibitem [{\citenamefont {Sheng}\ \emph {et~al.}(2005)\citenamefont {Sheng},
  \citenamefont {Sheng}, \citenamefont {Ting},\ and\ \citenamefont
  {Haldane}}]{PRL_2005}%
  \BibitemOpen
  \bibfield  {author} {\bibinfo {author} {\bibfnamefont {L.}~\bibnamefont
  {Sheng}}, \bibinfo {author} {\bibfnamefont {D.~N.}\ \bibnamefont {Sheng}},
  \bibinfo {author} {\bibfnamefont {C.~S.}\ \bibnamefont {Ting}},\ and\
  \bibinfo {author} {\bibfnamefont {F.~D.~M.}\ \bibnamefont {Haldane}},\
  }\bibfield  {title} {\bibinfo {title} {Nondissipative spin hall effect via
  quantized edge transport},\ }\href
  {https://doi.org/10.1103/PhysRevLett.95.136602} {\bibfield  {journal}
  {\bibinfo  {journal} {Phys. Rev. Lett.}\ }\textbf {\bibinfo {volume} {95}},\
  \bibinfo {pages} {136602} (\bibinfo {year} {2005})}\BibitemShut {NoStop}%
\bibitem [{\citenamefont {Hohenberg}\ and\ \citenamefont
  {Kohn}(1964)}]{kohan_dft}%
  \BibitemOpen
  \bibfield  {author} {\bibinfo {author} {\bibfnamefont {P.}~\bibnamefont
  {Hohenberg}}\ and\ \bibinfo {author} {\bibfnamefont {W.}~\bibnamefont
  {Kohn}},\ }\bibfield  {title} {\bibinfo {title} {Inhomogeneous electron
  gas},\ }\href {https://doi.org/10.1103/PhysRev.136.B864} {\bibfield
  {journal} {\bibinfo  {journal} {Phys. Rev.}\ }\textbf {\bibinfo {volume}
  {136}},\ \bibinfo {pages} {B864} (\bibinfo {year} {1964})}\BibitemShut
  {NoStop}%
\bibitem [{\citenamefont {Kresse}\ and\ \citenamefont
  {Furthm\"uller}(1996)}]{VASP}%
  \BibitemOpen
  \bibfield  {author} {\bibinfo {author} {\bibfnamefont {G.}~\bibnamefont
  {Kresse}}\ and\ \bibinfo {author} {\bibfnamefont {J.}~\bibnamefont
  {Furthm\"uller}},\ }\href {https://doi.org/10.1103/PhysRevB.54.11169}
  {\bibfield  {journal} {\bibinfo  {journal} {Phys. Rev. B}\ }\textbf {\bibinfo
  {volume} {54}},\ \bibinfo {pages} {11169} (\bibinfo {year}
  {1996})}\BibitemShut {NoStop}%
\bibitem [{\citenamefont {Perdew}\ \emph {et~al.}(1996)\citenamefont {Perdew},
  \citenamefont {Burke},\ and\ \citenamefont
  {Ernzerhof}}]{perdew1996generalized}%
  \BibitemOpen
  \bibfield  {author} {\bibinfo {author} {\bibfnamefont {J.~P.}\ \bibnamefont
  {Perdew}}, \bibinfo {author} {\bibfnamefont {K.}~\bibnamefont {Burke}},\ and\
  \bibinfo {author} {\bibfnamefont {M.}~\bibnamefont {Ernzerhof}},\ }\href
  {https://doi.org/10.1103/PhysRevLett.77.3865} {\bibfield  {journal} {\bibinfo
   {journal} {Phys. Rev. Lett.}\ }\textbf {\bibinfo {volume} {77}},\ \bibinfo
  {pages} {3865} (\bibinfo {year} {1996})}\BibitemShut {NoStop}%
\bibitem [{\citenamefont {Grimme}\ \emph {et~al.}(2010)\citenamefont {Grimme},
  \citenamefont {Antony}, \citenamefont {Ehrlich},\ and\ \citenamefont
  {Krieg}}]{Grimme2010}%
  \BibitemOpen
  \bibfield  {author} {\bibinfo {author} {\bibfnamefont {S.}~\bibnamefont
  {Grimme}}, \bibinfo {author} {\bibfnamefont {J.}~\bibnamefont {Antony}},
  \bibinfo {author} {\bibfnamefont {S.}~\bibnamefont {Ehrlich}},\ and\ \bibinfo
  {author} {\bibfnamefont {H.}~\bibnamefont {Krieg}},\ }\bibfield  {title}
  {\bibinfo {title} {A consistent and accurate ab initio parametrization of
  density functional dispersion correction ({DFT-D}) for the 94 elements
  {H-Pu}},\ }\href {https://doi.org/10.1063/1.3382344} {\bibfield  {journal}
  {\bibinfo  {journal} {J. Chem. Phys.}\ }\textbf {\bibinfo {volume} {132}},\
  \bibinfo {pages} {154104} (\bibinfo {year} {2010})}\BibitemShut {NoStop}%
\bibitem [{\citenamefont {Mostofi}\ \emph {et~al.}(2008)\citenamefont
  {Mostofi}, \citenamefont {Yates}, \citenamefont {Lee}, \citenamefont {Souza},
  \citenamefont {Vanderbilt},\ and\ \citenamefont
  {Marzari}}]{mostofi2008wannier90}%
  \BibitemOpen
  \bibfield  {author} {\bibinfo {author} {\bibfnamefont {A.~A.}\ \bibnamefont
  {Mostofi}}, \bibinfo {author} {\bibfnamefont {J.~R.}\ \bibnamefont {Yates}},
  \bibinfo {author} {\bibfnamefont {Y.-S.}\ \bibnamefont {Lee}}, \bibinfo
  {author} {\bibfnamefont {I.}~\bibnamefont {Souza}}, \bibinfo {author}
  {\bibfnamefont {D.}~\bibnamefont {Vanderbilt}},\ and\ \bibinfo {author}
  {\bibfnamefont {N.}~\bibnamefont {Marzari}},\ }\href
  {https://doi.org/https://doi.org/10.1016/j.cpc.2007.11.016} {\bibfield
  {journal} {\bibinfo  {journal} {Comput. Phys. Commun.}\ }\textbf {\bibinfo
  {volume} {178}},\ \bibinfo {pages} {685} (\bibinfo {year}
  {2008})}\BibitemShut {NoStop}%
\bibitem [{\citenamefont {Wu}\ \emph {et~al.}(2018)\citenamefont {Wu},
  \citenamefont {Zhang}, \citenamefont {Song}, \citenamefont {Troyer},\ and\
  \citenamefont {Soluyanov}}]{wu2018wanniertools}%
  \BibitemOpen
  \bibfield  {author} {\bibinfo {author} {\bibfnamefont {Q.}~\bibnamefont
  {Wu}}, \bibinfo {author} {\bibfnamefont {S.}~\bibnamefont {Zhang}}, \bibinfo
  {author} {\bibfnamefont {H.-F.}\ \bibnamefont {Song}}, \bibinfo {author}
  {\bibfnamefont {M.}~\bibnamefont {Troyer}},\ and\ \bibinfo {author}
  {\bibfnamefont {A.~A.}\ \bibnamefont {Soluyanov}},\ }\href
  {https://doi.org/https://doi.org/10.1016/j.cpc.2017.09.033} {\bibfield
  {journal} {\bibinfo  {journal} {Comput. Phys. Commun.}\ }\textbf {\bibinfo
  {volume} {224}},\ \bibinfo {pages} {405} (\bibinfo {year}
  {2018})}\BibitemShut {NoStop}%
\bibitem [{\citenamefont {Sancho}\ \emph {et~al.}(1985)\citenamefont {Sancho},
  \citenamefont {Sancho}, \citenamefont {Sancho},\ and\ \citenamefont
  {Rubio}}]{Greenwanniertools}%
  \BibitemOpen
  \bibfield  {author} {\bibinfo {author} {\bibfnamefont {M.~P.~L.}\
  \bibnamefont {Sancho}}, \bibinfo {author} {\bibfnamefont {J.~M.~L.}\
  \bibnamefont {Sancho}}, \bibinfo {author} {\bibfnamefont {J.~M.~L.}\
  \bibnamefont {Sancho}},\ and\ \bibinfo {author} {\bibfnamefont
  {J.}~\bibnamefont {Rubio}},\ }\href
  {https://doi.org/10.1088/0305-4608/15/4/009} {\bibfield  {journal} {\bibinfo
  {journal} {J. Phys. F: Met. Phys}\ }\textbf {\bibinfo {volume} {15}},\
  \bibinfo {pages} {851} (\bibinfo {year} {1985})}\BibitemShut {NoStop}%
\bibitem [{\citenamefont {Gao}\ \emph {et~al.}(2021)\citenamefont {Gao},
  \citenamefont {Wu}, \citenamefont {Persson},\ and\ \citenamefont
  {Wang}}]{irvsp}%
  \BibitemOpen
  \bibfield  {author} {\bibinfo {author} {\bibfnamefont {J.}~\bibnamefont
  {Gao}}, \bibinfo {author} {\bibfnamefont {Q.}~\bibnamefont {Wu}}, \bibinfo
  {author} {\bibfnamefont {C.}~\bibnamefont {Persson}},\ and\ \bibinfo {author}
  {\bibfnamefont {Z.}~\bibnamefont {Wang}},\ }\bibfield  {title} {\bibinfo
  {title} {Irvsp: To obtain irreducible representations of electronic states in
  the {VASP}},\ }\href {https://doi.org/10.1016/j.cpc.2020.107760} {\bibfield
  {journal} {\bibinfo  {journal} {Comput. Phys. Commun.}\ }\textbf {\bibinfo
  {volume} {261}},\ \bibinfo {pages} {107760} (\bibinfo {year}
  {2021})}\BibitemShut {NoStop}%
\bibitem [{\citenamefont {Aroyo}\ \emph
  {et~al.}(2006{\natexlab{a}})\citenamefont {Aroyo}, \citenamefont
  {Perez-Mato}, \citenamefont {Capillas}, \citenamefont {Kroumova},
  \citenamefont {Ivantchev}, \citenamefont {Madariaga}, \citenamefont {Kirov},\
  and\ \citenamefont {Wondratschek}}]{bilbao_1}%
  \BibitemOpen
  \bibfield  {author} {\bibinfo {author} {\bibfnamefont {M.~I.}\ \bibnamefont
  {Aroyo}}, \bibinfo {author} {\bibfnamefont {J.~M.}\ \bibnamefont
  {Perez-Mato}}, \bibinfo {author} {\bibfnamefont {C.}~\bibnamefont
  {Capillas}}, \bibinfo {author} {\bibfnamefont {E.}~\bibnamefont {Kroumova}},
  \bibinfo {author} {\bibfnamefont {S.}~\bibnamefont {Ivantchev}}, \bibinfo
  {author} {\bibfnamefont {G.}~\bibnamefont {Madariaga}}, \bibinfo {author}
  {\bibfnamefont {A.}~\bibnamefont {Kirov}},\ and\ \bibinfo {author}
  {\bibfnamefont {H.}~\bibnamefont {Wondratschek}},\ }\bibfield  {title}
  {\bibinfo {title} {Bilbao crystallographic server: I. databases and
  crystallographic computing programs},\ }\href
  {https://doi.org/10.1524/zkri.2006.221.1.15} {\bibfield  {journal} {\bibinfo
  {journal} {Z. Kristallogr. Cryst. Mater.}\ }\textbf {\bibinfo {volume}
  {221}},\ \bibinfo {pages} {15} (\bibinfo {year}
  {2006}{\natexlab{a}})}\BibitemShut {NoStop}%
\bibitem [{\citenamefont {Aroyo}\ \emph
  {et~al.}(2006{\natexlab{b}})\citenamefont {Aroyo}, \citenamefont {Kirov},
  \citenamefont {Capillas}, \citenamefont {Perez-Mato},\ and\ \citenamefont
  {Wondratschek}}]{bilbao_2}%
  \BibitemOpen
  \bibfield  {author} {\bibinfo {author} {\bibfnamefont {M.~I.}\ \bibnamefont
  {Aroyo}}, \bibinfo {author} {\bibfnamefont {A.}~\bibnamefont {Kirov}},
  \bibinfo {author} {\bibfnamefont {C.}~\bibnamefont {Capillas}}, \bibinfo
  {author} {\bibfnamefont {J.~M.}\ \bibnamefont {Perez-Mato}},\ and\ \bibinfo
  {author} {\bibfnamefont {H.}~\bibnamefont {Wondratschek}},\ }\bibfield
  {title} {\bibinfo {title} {Bilbao crystallographic server. {II}.
  representations of crystallographic point groups and space groups},\ }\href
  {https://doi.org/10.1107/s0108767305040286} {\bibfield  {journal} {\bibinfo
  {journal} {Acta Crystallographica Section A Foundations of Crystallography}\
  }\textbf {\bibinfo {volume} {62}},\ \bibinfo {pages} {115} (\bibinfo {year}
  {2006}{\natexlab{b}})}\BibitemShut {NoStop}%
\bibitem [{\citenamefont {Aroyo}\ \emph {et~al.}(2011)\citenamefont {Aroyo},
  \citenamefont {Perez-Mato}, \citenamefont {Orobengoa}, \citenamefont {Tasci},
  \citenamefont {de~la Flor},\ and\ \citenamefont {Kirov}}]{Bilbao_online}%
  \BibitemOpen
  \bibfield  {author} {\bibinfo {author} {\bibfnamefont {M.~I.}\ \bibnamefont
  {Aroyo}}, \bibinfo {author} {\bibfnamefont {J.~M.}\ \bibnamefont
  {Perez-Mato}}, \bibinfo {author} {\bibfnamefont {D.}~\bibnamefont
  {Orobengoa}}, \bibinfo {author} {\bibfnamefont {E.}~\bibnamefont {Tasci}},
  \bibinfo {author} {\bibfnamefont {G.}~\bibnamefont {de~la Flor}},\ and\
  \bibinfo {author} {\bibfnamefont {A.}~\bibnamefont {Kirov}},\ }\bibfield
  {title} {\bibinfo {title} {Crystallography online: Bilbao crystallographic
  server},\ }\href {https://www.cryst.ehu.es/} {\bibfield  {journal} {\bibinfo
  {journal} {Bulg. Chem. Commun.}\ }\textbf {\bibinfo {volume} {43}},\ \bibinfo
  {pages} {183} (\bibinfo {year} {2011})}\BibitemShut {NoStop}%
\end{thebibliography}%

\renewcommand{\thefigure}{S\arabic{figure}}
\renewcommand{\thesection}{S-\Roman{section}}
\renewcommand{\thetable}{S\Roman{table}}

\setcounter{figure}{0}
\setcounter{section}{0}
\clearpage
\title{--Supplemental Material-- \\Atomically thin obstructed atomic insulators with robust edge modes and quantized spin Hall effect}
\maketitle
\onecolumngrid
\section{Methods and crystal structure}~
Electronic structure calculations were performed within the density functional theory framework with projector augmented-wave potentials using the Vienna \textit{ab-initio} simulation package (VASP)~\cite{kohan_dft,VASP}. A kinetic energy cut-off of 420 eV for the plane-wave basis set and Gaussian smearing with a smearing width of 50 meV to define the partial occupancies for each wave function were used. The exchange-correlation effects were treated within the generalized gradient approximation (GGA) with van der Waals corrections, and the SOC was added self-consistently to include the relativistic effects~\cite{perdew1996generalized,Grimme2010}. We used $\Gamma-$centered $19 \times 21 \times 1$ and $16 \times 1\times 1$ $k$-meshes to sample the 2D and 1D BZs. The monolayer structures were constructed by adding a vacuum larger than 15 {\AA} to avoid interactions between the periodically repeated images. The internal parameters were relaxed until the forces on each atom were less than $10^{-3}$ eV{\AA}$^{-1}$ and electronic energy minimization tolerance was set to $10^{-6}$ eV. We also constructed material-specific tight-binding model Hamiltonians from the atom-centered Wannier functions to calculate topological properties~\cite{mostofi2008wannier90, wu2018wanniertools, Greenwanniertools}.

\begin{figure}[b]
\includegraphics[width=0.85\linewidth]{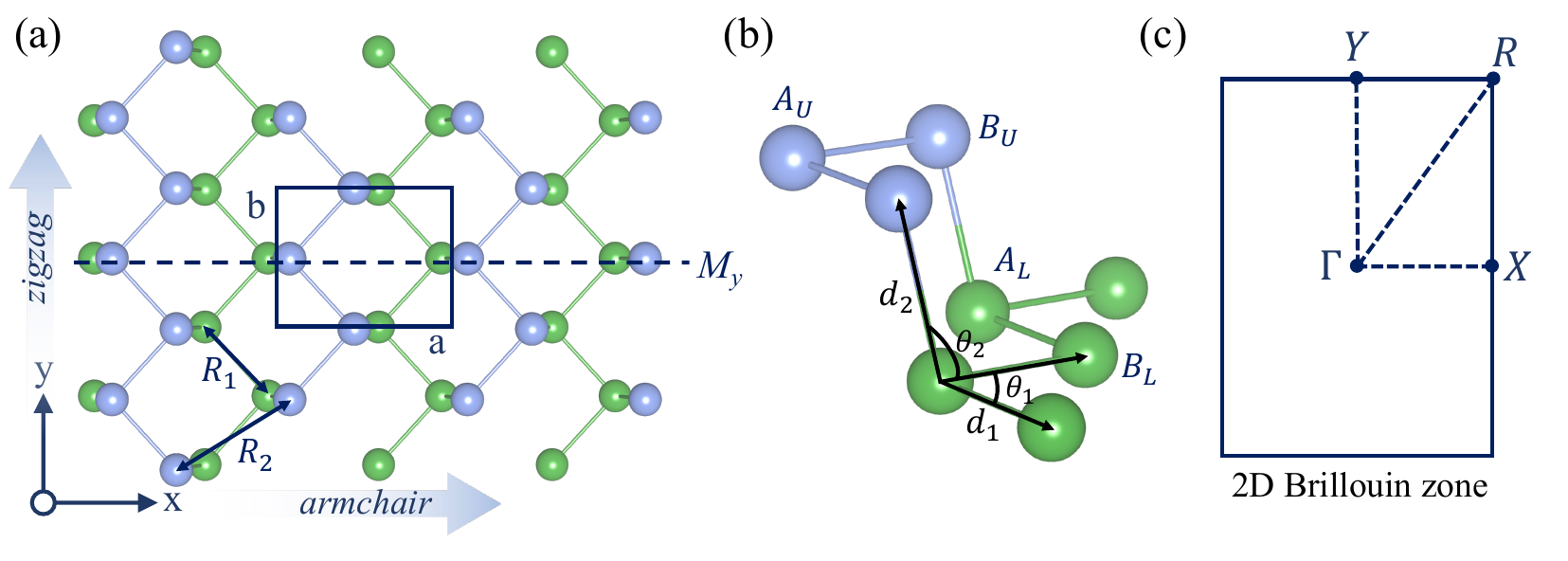}
\caption{ (a) Top and (b) side view of the monolayer crystal structure of puckered lattice. The green balls represent the atoms at occupied Wyckoff position $4h$ as discussed in the main text. Blue rectangle marks the primitive unit cell with lattice parameters $a$ and $b$ along armchair ($x-$axis) and zigzag ($y-$axis) directions, respectively. The bond angles $\theta_1$ and $\theta_2$ are marked in (b). (c) Associated two-dimensional Brillouin zone with high symmetry points \{$\Gamma$, X, R, Y\}.}
\label{structure}
\end{figure}

The monolayer puckered lattice structure of phosphorene or other group-Va monolayers is shown in Fig.~\ref{structure}(a). This structure consists of four atoms arranged in two atomic planes, named A$_U$, A$_L$, B$_U$, and B$_L$, with U and L indicating the upper and lower planes, respectively (Fig.~\ref{structure}(b)). Within each plane, the atoms form zigzag chains oriented along the $y$-axis, whereas, the co-planar zigzag chains are connected via an out-of-plane bond, resulting in an armchair structure along the $x-$axis. The corresponding primitive unit cell is rectangular, characterized by lattice parameters $a$ and $b$ along the $x-$ and $y-$axes, respectively. Fig.~\ref{structure}(b) provides a side view of the monolayer lattice, highlighting the two bond angles $\theta_1$ and $\theta_2$ as well as the interatomic distances $d_1$ and $d_2$. $R_1$, $R_2$, and $d_2$ in Figs.~\ref{structure}(a)-(b) define the hopping parameters $t_1$, $t_2$, and $t_\perp$, respectively, as used in the model Hamiltonian given in the main text.

We present the optimized structural parameters of group-Va monolayers using GGA exchange-correlation functional with van der Waals correction in Table~\ref{optimized_geometry}. The lattice constants and in-plane and out-of-plane bond lengths ($d_1$ and $d_2$) increase from P to Sb. However, the in-plane and out-of-plane bond angles change their order as one moves from P to Sb. Specifically, $\theta_2 > \theta_1$ for both P and As. The bond angle order is reversed for Sb with $ \theta_1> \theta_2$. As discussed in the main text, the emergence of an inverted topological phase in these monolayers is strongly correlated with a change in the bond angles. In Fig.~\ref{structure}(c), we present two-dimensional Brillouin zone (BZ) with four high symmetry points $\Gamma=(0, 0)$, $X=(\frac{\pi}{a}, 0)$, $Y=(0, \frac{\pi}{b})$ and $R=(\frac{\pi}{a}, \frac{\pi}{b})$ marked. 

\begin{table}[h]
\caption{Optimized lattice parameter of monolayer P, As, and Sb with puckered lattice (space group $Pmna$, $\# 53$) obtained using GGA with van der Waals correction. $a$, $b$ are the lattice parameters along the $x-$axis and $y-$axis, $d_1$, $d_2$ are in-plane and out-of-plane bond lengths, and $\theta_1$, $\theta_2$ are co-planar and non-co-planer bond angles (see Fig.~\ref{structure}(b)).}
\begin{tabular}{l c c c c c c c c }
\hline\hline
			& &  $a$ (\AA)		&~~ $b$ (\AA)		&~~ d$_{1} $(\AA)	&~~  d$_{2}$(\AA)  	&~~ $\theta_{1}$(in degrees)	&~~ $\theta_{2}$(in degrees)      \\
\hline
P                           	& &  4.588  		&~~ 3.296       		&~~ 2.218    		& ~~ 2.260		&~~95.97			&~~103.87			\\
As                           	& &  4.679  		&~~ 3.700  		&~~ 2.509   		&~~  2.493  		&~~94.99			&~~100.05 			\\
Sb                                 &  &  4.779 		&~~ 4.373   		&~~ 2.941   		& ~~ 2.855		&~~96.04			&~~95.67	  		\\
\hline 
\end{tabular}
\label{optimized_geometry}
\end{table}

\section{Graphene vs nontrivial inverted A\lowercase{s} bands}

\begin{figure}[ht]
\includegraphics[width=0.85\linewidth]{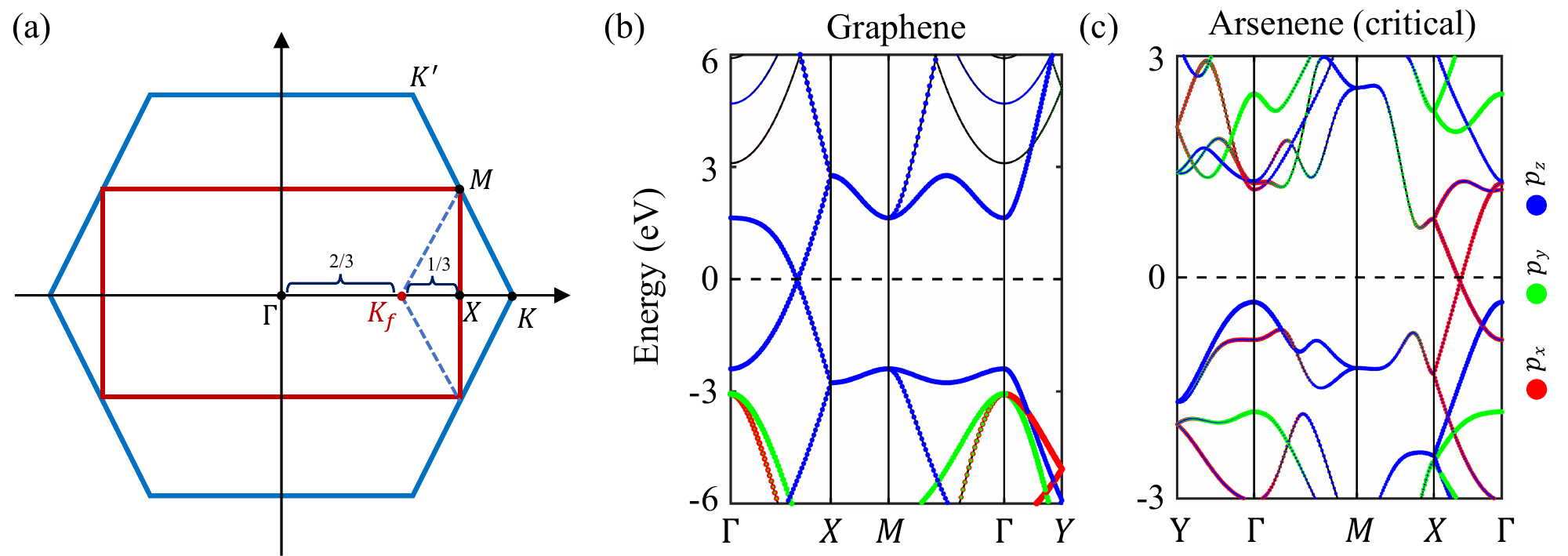}
\caption{(a) Folding of hexagonal BZ onto a puckered hexagonal (rectangular) BZ. $K/K^\prime$ valleys of hexagonal BZ folds to $K_f$ on the $\Gamma-X$ line. Calculated band structure of (b) graphene and (c) strained As (at critical point). Red, green, and blue represent the $p_x$, $p_y$, and $p_z$ orbital contribution.}
\label{bands}
\end{figure}

We now compare the characteristics of graphene's electronic bands on a folded rectangular Brillouin zone (BZ) with those at the critical point of strained arsenene (As) when $b' = 1.12b$.  Figure~\ref{bands}(a) shows the folding of the hexagonal BZ of graphene into a rectangular BZ that arises due to the lattice puckering. The $K$ point of the hexagonal BZ projects onto a point $K_f$ in the rectangular BZ, situated at a distance of ${2\over 3} \times {\Gamma-X}$. Consequently, the $K(K')$ valley of the folded graphene bands appears at $K_f$ within the rectangular BZ, where the Dirac bands are projected (see Fig.~\ref{bands}(b)].

In contrast, at the critical point of strained As with $b^\prime = 1.12b $, the Dirac nodes are located at $0.62 \times {\Gamma-X}$, which is $\sim 6\%$ away from folded $K/K^\prime$ points of the hexagonal BZ. Moreover, the Dirac nodes in graphene originate from out-of-plane $p_z$ orbitals, whereas those in As or Sb are derived from in-plane $p_x$ orbitals. These results suggest a direct one-to-one mapping between the critical point Dirac nodes of Sb or strained As and the $K/K'$ valleys of hexagonal graphene may not be expected.

\section{Doping dependence band structure of A\lowercase{s$_{1-x}$}S\lowercase{b$_{x}$}}

\begin{figure}[h]
\includegraphics[width=0.85\linewidth]{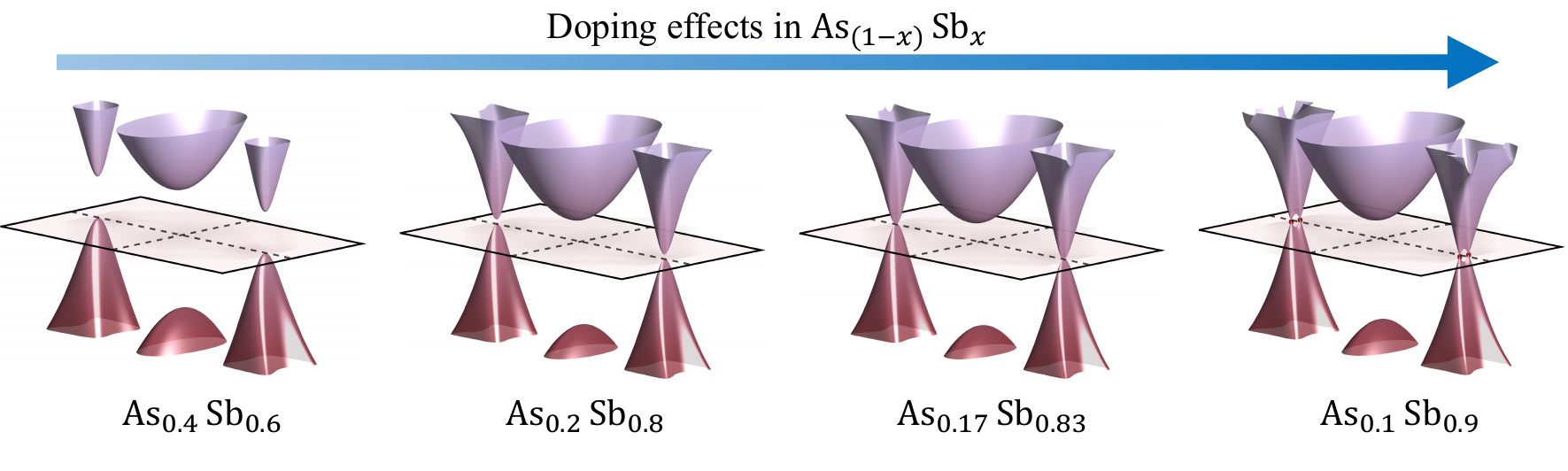}
\caption{Calculated band structure of As$_{1-x}$Sb$_x$ as a function of $x$ without SOC. The unpinned Dirac cones are marked with circles for $x=0.9$.}
\label{tpt_strain}
\end{figure}

A nontrivial topological state characterized by band inversions at generic $k$-points in group-Va monolayers can also be realized by varying the concentration of Sb in As$_{1-x}$Sb$_x$. To model the electronic structure of As$_{1-x}$Sb$_x$, we employ the virtual crystal approximation. In Fig.~\ref{tpt_strain}, we show the evolution of band structures of As$_{1-x}$Sb$_x$ as a function of Sb concentration without SOC. When $x=0.83$, the band gap closes at two generic $k$ points on the $\Gamma-X$ line. As we increase $x>0.83$, each critical Dirac node split into a pair of Dirac nodes located away from the $\Gamma-X$ line. The emergence of generic point Dirac nodes and their hybridization under SOC to realize quantized spin Hall effect is similar to Sb or strained As as discussed in the main text.

\section{Edge band structure of phosphorene with SOC}

\begin{figure}[h]
\includegraphics[width=0.85\linewidth]{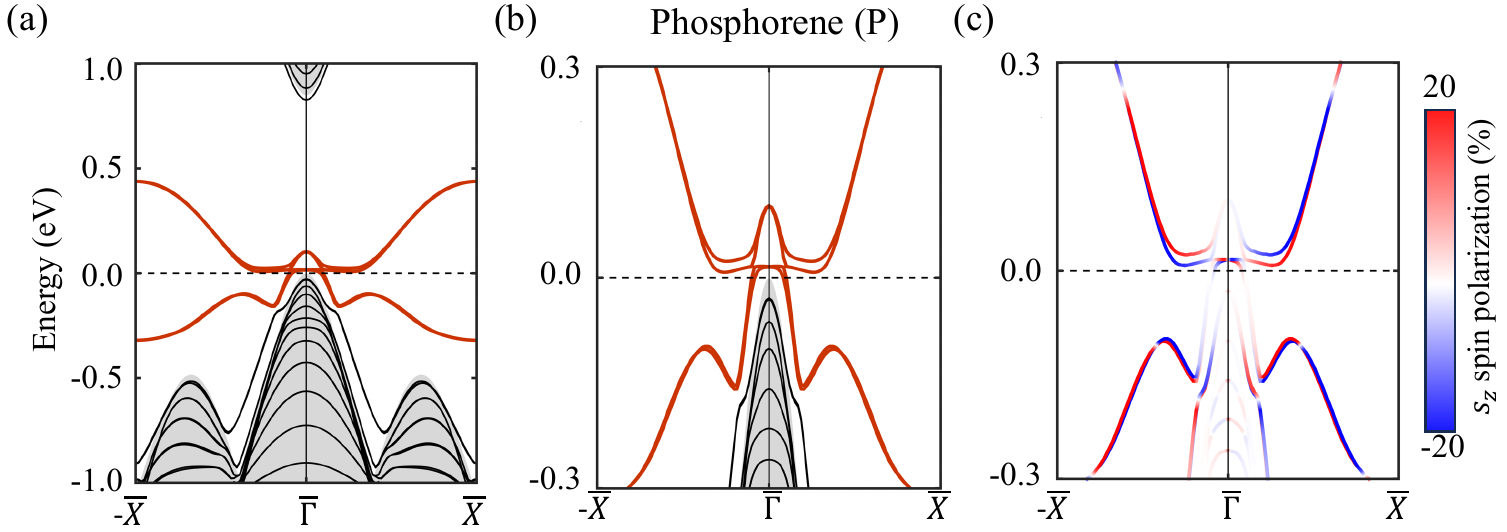}
\caption{(a) Calculated edge band structure of phosphorene with spin-orbit coupling (SOC). Closeup of bands near the $\overline{\Gamma}$ point (b) without and (c) with spin-texture. Blue and red colors indicate spin-up and spin-down channels of $S_z$.}
\label{edgebands}
\end{figure}

Figure~\ref{edgebands} illustrates the calculated armchair edge band structure of phosphorene with spin-orbit coupling (SOC). The inclusion of SOC causes spin-splitting of the edge bands, similar to what is observed in arsenene, as discussed in the main text. However, due to the relatively weak SOC in phosphorus atoms, the resulting spin-splitting in the bands is small. It is also important to note that the spin polarization in these edge states (OESs) near the $\overline{\Gamma}$ point is reduced due to the mixing of various bands with the OESs.

\newpage
\section{Band Representations}~
We present the atom-induced band representations (aBRs) of occupied $4h$ Wyckoff position for P, As, and Sb in Table.~\ref{aBR}. $4h$ Wyckoff position gives rise to two possible aBRs, A$^\prime$ and A$^{\prime \prime}$ that are induced from basis $s$, $p_x/p_y$, and $p_z$ of atom-occupied Wyckoff position $4h$. We evaluate the band representations (BRs) by mapping the elementary band representations (eBRs) with the irreducible representations (IRs) at the high-symmetry points. As shown in Fig.~1(b) of the main text, the lower four bands with $s$ orbital character are induced from atom-occupied Wyckoff position with band representation A$^\prime$@4h. The remaining six occupied bands are induced from atom-unoccupied Wyckoff positions $2c$ and $4g$ with band representation A$_g$@{2c} and A@{4g}. We use space-group standardized structure files to evaluate BRs and IRs using \href{https://github.com/zjwang11/IRVSP}{IRVSP}, \href{https://tm.iphy.ac.cn/UnconvMat.html}{pos2aBR}, and \href{https://www.cryst.ehu.es/}{Bilbao crystallographic server}~\cite{irvsp,bilbao_1, bilbao_2,Bilbao_online}.
 
\begin{table}[h]
\caption{Atom-induced band representations (aBRs) of P, As, and Sb in space group $Pmna$. $q$ gives Wyckoff position, $\rho$@q represents the band representation from $q$, and IRs ($\rho$)  gives irreducible representations (IRs) associated with  $\rho$.}
\begin{tabular}{l c c c r c c }
\\
\hline \hline 
	 ~~Atom      &~~WP(q)         &~~Site Symm.        &~~Configuration               &~~IRs($\rho$)                                    & ~~aBRs($\rho$@q)        \\ 
\hline  
              ~~P          &~~ 4$h$              &~~$m$                    &~~3s$^2$ 3p$^3$            &~~ $s$ :~~A$^\prime$                             & A$^\prime$@4h                   \\
                              &                            &                              &                                         &~~ $p_x$, $p_y$ :~~A$^\prime$             & A$^\prime$@4h                       \\
                              &                            &                              &                                         &~~ $p_z$ :~~A$^{\prime \prime}$            & A$^{\prime \prime}$@4h        \\
\hline 
              ~~As        &~~ 4$h$               &~~ $m$                 &~~4s$^2$ 4p$^3$            &~~ $s$ :~~A$^\prime$                             & A$^\prime$@4h               \\
                              &                            &                             &                                         &~~ $p_x$, $p_y$ :~~A$^\prime$             & A$^\prime$@4h                       \\
                              &                            &                             &                                         &~~ $p_z$ :~~A$^{\prime \prime}$            & A$^{\prime \prime}$@4h      \\
\hline
              ~~Sb       &~~ 4$h$               &~~ $m$                  &~~5s$^2$ 5p$^3$            &~~ $s$ :~~A$^\prime$                             & A$^\prime$@4h               \\
                              &                            &                              &                                         &~~ $p_x$, $p_y$ :~~A$^\prime$             & A$^\prime$@4h                       \\
                              &                           &                              &                                          &~~ $p_z$ :~~A$^{\prime \prime}$            & A$^{\prime \prime}$@4h      \\
\hline
\end{tabular}
\label{aBR}
\end{table}

\begin{table}[p]
\renewcommand*{\arraystretch}{1.2}
\setlength{\tabcolsep}{4pt}
\caption{Irreducible representations (IRs) and band representations (BRs) associated with ten occupied bands of monolayer P, As, and Sb without spin-orbit coupling (SOC). BRs from the atom-unoccupied Wyckoff positions ($A@4g$ and $A_g@2c$) are shown in bold. IRs at various high-symmetry points are given in increasing order of band energy. $A_i^{\pm}(n)$ marks a IR with parity $\pm$ and degeneracy $n$.}
\begin{tabular}{c c c c c }
\hline
				         &\multicolumn{4}{c}{Phosphorene (P)}\\
\hline 
	                                               &~~$\Gamma$                 &~~$X$                         &~~$R$                         &~~$Y$    \\ 	      
	                                         
\hline         
	\multirow{10}{3em}{Bands}    &~~$\Gamma_1^+(1)$      &~~$X_1 (2)$  	&~~$R_1^-(2)$               &~~$Y_1 (2)$ \\   
				         &~~$\Gamma_4^-(1)$      &~~$X_1 (2)$  	            	&~~$R_1^+(2)$               &~~$Y_2 (2)$ \\ 
				         &~~$\Gamma_3^+(1)$      &~~$X_1 (2)$  	&~~$R_1^-(2)$               &~~$Y_1 (2)$ \\ 
				         &~~$\Gamma_1^+(1)$      &~~$X_2 (2)$  	&~~$R_1^+(2)$               &~~$Y_2 (2)$ \\ 
				         &~~$\Gamma_2^-(1)$      &~~$X_1 (2)$    	&~~$R_1^-(2)$               &~~$Y_1 (2)$ \\
				         &~~$\Gamma_4^-(1)$       &           	            &                                      &                   \\
				         &~~$\Gamma_1^-(1)$       &           	            &                                      &                   \\
				         &~~$\Gamma_4^+(1)$      &           	            &                                      &                   \\
				         &~~$\Gamma_1^+(1)$      &           	            &                                      &                   \\
				         &~~$\Gamma_3^+(1)$      &           	            &                                      &                   \\
				         
\hline
				         &\multicolumn{4}{c}{Arsenene (As)}\\
\hline 

	\multirow{10}{3em}{Bands}    &~~$\Gamma_1^+(1)$      &~~$X_1 (2)$  	&~~$R_1^-(2)$               &~~$Y_1 (2)$ \\   
				         &~~$\Gamma_4^-(1)$      &~~$X_1 (2)$  	            	&~~$R_1^+(2)$               &~~$Y_2 (2)$ \\ 
				         &~~$\Gamma_3^+(1)$      &~~$X_1 (2)$  	&~~$R_1^+(2)$               &~~$Y_1 (2)$ \\ 
				         &~~$\Gamma_2^-(1)$      &~~$X_2 (2)$  	             &~~$R_1^-(2)$               &~~$Y_2 (2)$ \\ 
				         &~~$\Gamma_1^+(1)$      &~~$X_1 (2)$  	&~~$R_1^-(2)$               &~~$Y_1 (2)$ \\
				         &~~$\Gamma_4^-(1)$       &           	            &                                      &                   \\
				         &~~$\Gamma_1^-(1)$       &           	            &                                      &                   \\
				         &~~$\Gamma_1^+(1)$      &           	            &                                      &                   \\
				         &~~$\Gamma_4^+(1)$      &           	            &                                      &                   \\
				         &~~$\Gamma_3^+(1)$      &           	            &                                      &                   \\
\hline
				         &\multicolumn{4}{c}{Antimony (Sb)}\\
\hline 

	\multirow{10}{3em}{Bands}    &~~$\Gamma_1^+(1)$      &~~$X_1 (2)$  	&~~$R_1^-(2)$               &~~$Y_1 (2)$ \\   
				         &~~$\Gamma_4^-(1)$      &~~$X_1 (2)$  	            	&~~$R_1^+(2)$               &~~$Y_2 (2)$ \\ 
				         &~~$\Gamma_3^+(1)$      &~~$X_1 (2)$  	&~~$R_1^+(2)$               &~~$Y_1 (2)$ \\ 
				         &~~$\Gamma_2^-(1)$      &~~$X_2 (2)$  	             &~~$R_1^-(2)$               &~~$Y_2 (2)$ \\ 
				         &~~$\Gamma_1^+(1)$      &~~$X_1 (2)$  	&~~$R_1^-(2)$               &~~$Y_1 (2)$ \\
				         &~~$\Gamma_4^-(1)$       &           	            &                                      &                   \\
				         &~~$\Gamma_1^-(1)$       &           	            &                                      &                   \\
				         &~~$\Gamma_1^+(1)$      &           	            &                                      &                   \\
				         &~~$\Gamma_4^+(1)$      &           	            &                                      &                   \\
				         &~~$\Gamma_3^+(1)$      &           	            &                                      &                   \\
\hline
	 \multicolumn{1}{l}{BRs}\\
\hline
	\multirow{1}{5em}{A$^\prime$@{4h}}    
				         &~~$\Gamma_1^+(1)\oplus \Gamma_2^-(1)\oplus \Gamma_3^+(1)\oplus \Gamma_4^-(1)$
				         &~~$2X_1(2)$               
				         &~~$R_1^+(2) \oplus R_1^-(2)$
				         &~~$Y_1(2) \oplus Y_2(2)$ \\   
\hline 
	\multirow{1}{5em}{\textbf{A@{4g}}}    
				         &~~$\mathbf{\Gamma_1^+(1)\oplus \Gamma_1^-(1)\oplus \Gamma_4^+(1)\oplus \Gamma_4^-(1)}$
				         &~~$\mathbf{X_1(2)\oplus X_2(2)}$               
				         &~~$\mathbf{R_1^+(2) \oplus R_1^-(2)}$
				          &~~$\mathbf{Y_1(2) \oplus  Y_2(2)}$ \\    
	\multirow{1}{5em}{\textbf{A$_g$@{2c}}} 
				         &~~$\mathbf{\Gamma_1^+(1)\oplus \Gamma_3^+(1)}$
				         &~~$\mathbf{X_1(2)}$               
				         &~~$\mathbf{R_1^-(2)}$
				         &~~$\mathbf{Y_1(2)}$ \\  
\hline                                                                                                                                                                                                                                                                                                 
\end{tabular}
\label{band_representation}
\end{table}

\newpage
\section{Spin-texture of edge states}
\begin{figure}[h]
\centering
\includegraphics[width=0.85\linewidth]{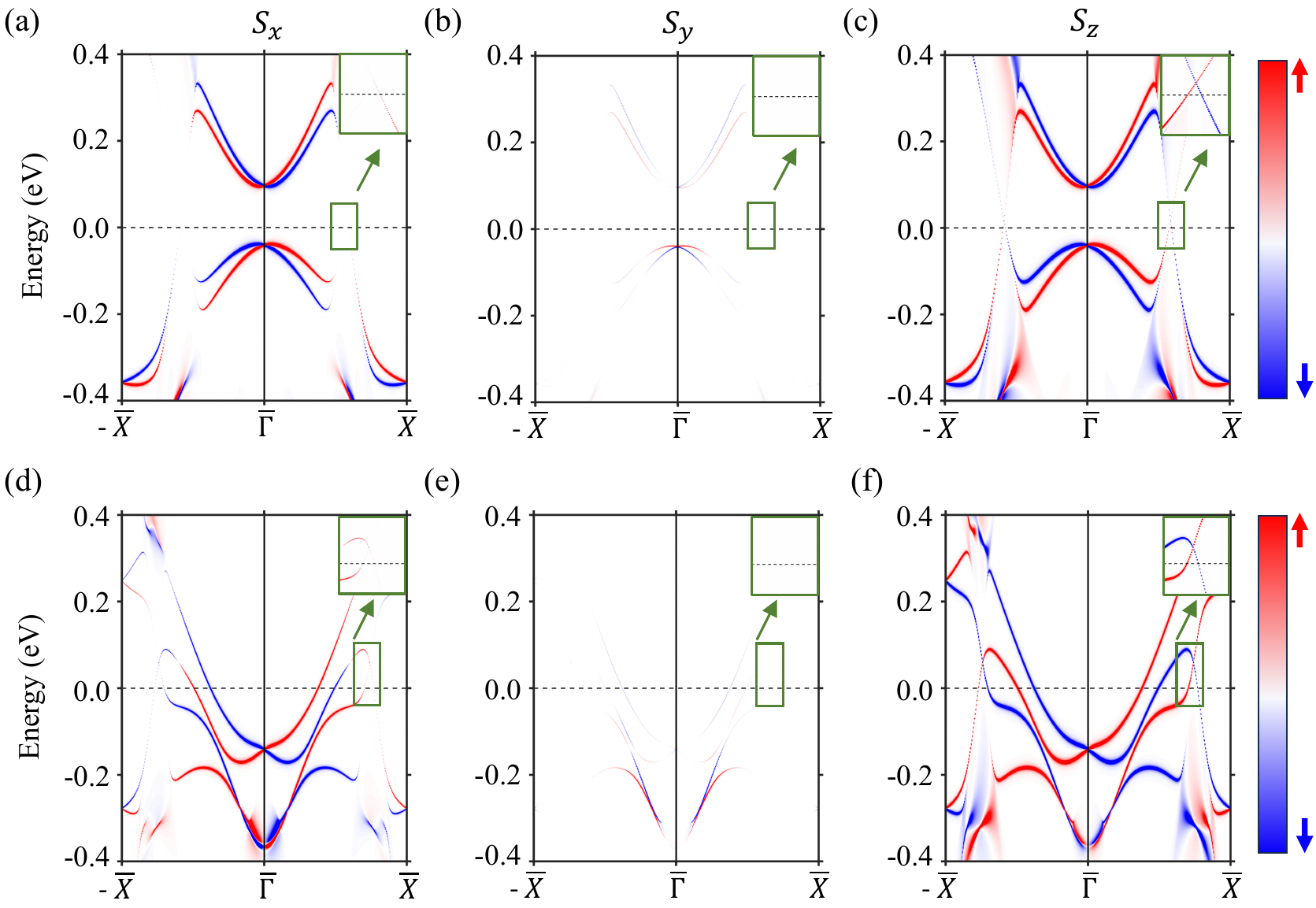}
\caption{Spin-texture of edge states of (a-c) nontrivial inverted As ($b'=1.14b$) and (d-f) Sb. Red and blue colors represent the up and down channels of a spin component. The insets highlight the spin polarization of the edge bands in proximity to the bulk inverted region. The significant contribution to the spin texture primarily arises from the $S_z$ spin component, indicating a spin U(1) quasi-symmetry.}
 \label{spin_pol}
\end{figure}

We present the spin texture of the edge states of strained As and Sb in Fig.~\ref{spin_pol}. The presence of time-reversal symmetry and $\mathcal{C}_{2y}$ rotational symmetry along the armchair edge enforces $S_y=0$. In Fig.~\ref{spin_pol}, we present $S_x$, $S_y$, and $S_z$ spin resolved edge spectrum of nontrivial inverted As (As$_{NTI}$) and Sb along -$\overline{X}-\overline{\Gamma}-\overline{X}$ direction. The $S_y$ spin component is nearly zero, whereas $S_x$ and $S_z$ components are nonzero as dictated by the edge symmetries. Moreover, the edge states retain only $S_z$ spin component in the bulk band inversion region with vanishing $S_x$. Such a spin-texture of edge states ensures a spin U(1) quasi-symmetry with nearly $z$ polarized spin~\cite{QuasiSpin}. This is in accord with the SOC Hamiltonian presented for these monolayers in our main text and reported in Ref.~\cite{Unpinned_Dirac}. 

\end{document}